\documentclass{article}
\usepackage{authblk}

\usepackage{setspace}
\usepackage{amsmath,amssymb,amsbsy}
\usepackage{booktabs,hyperref,xcolor,gensymb}
\usepackage[a4paper, margin=1in]{geometry}
\usepackage[authoryear]{natbib}
\usepackage{bm,rotating}
\usepackage{subfigure}
\usepackage{multicol,multirow, array}

\usepackage{tikz}
\usetikzlibrary{trees,matrix,arrows,fit,positioning,shapes}

\newcolumntype{C}[1]{>{\centering\let\newline\\\arraybackslash\hspace{0pt}}m{#1}}

\title{Quantifying Responsibility with Probabilistic Causation --- The Case of Climate Action}

\author[a,b,*]{Sarah Hiller}
\author[b]{Jobst Heitzig}
\affil[a]{Free University Berlin, Institute for Mathematics, Arnimallee 3, 14195 Berlin, Germany}
\affil[b]{Potsdam Institute for Climate Impact Research, PO Box 60 12 03, 14412 Potsdam, Germany}
\affil[*]{Corresponding author. Email: sarah.hiller@fu-berlin.de}
\date{}

\def\eqdef{:=}

\def\rho{\varrho}
\def\ge{\geqslant}
\def\geq{\geqslant}
\def\le{\leqslant}

\def\ignore#1{}

\def\T{\mathcal{T}}
\def\R{\mathcal{R}}
\def\L{\mathcal{L}}
\def\eps{\varepsilon}
\def\N2{\frac{N}{2}}

\begin{document}
\onehalfspacing

\maketitle

\begin{abstract} 
    Many real-world situations of ethical and economic relevance, such as collective (in)action with respect to the climate crisis, involve not only diverse agents whose decisions interact in complicated ways, but also various forms of uncertainty, including both quantifiable risk and unquantifiable ambiguity. 
    In such cases, an assessment of moral responsibility for ethically undesired outcomes or of the responsibility to avoid these is challenging and prone to the risk of under- or overdetermination.
    In contrast to existing approaches that employ notions of causation based on combinations of necessity and sufficiency,
    or certain logics that focus on a binary classification of `responsible' vs `not responsible', 
    we present a set of quantitative metrics that assess responsibility degrees in units of probability.
    To this end, we adapt extensive-form game trees as the framework for representing decision scenarios and evaluate the proposed responsibility functions based on the correct representation of a set of analytically assessed paradigmatic example scenarios.
    We test the best performing metrics on a reduced representation of a real-world decision scenario and are able to compute meaningful responsibility scores.
\end{abstract}


\section{Introduction} \label{sec:intro}

The current climate crisis and its associated effects constitute one of the essential challenges for humanity and collective decision making in the upcoming years. An increase of greenhouse gas (GHG)\footnote{Prominently CO$_2$, but also methane, nitrous oxide and others.} concentrations in the atmosphere attributable to human activity leads to a warming of Earth's surface temperature by reducing the fraction of incoming solar radiation that is diffused back into space. An elevated mean Earth surface temperature is however not a priori something reprehensible. Rather, it is the resultant effects that carry enormous dangers. Among these are the increased risk of extreme weather events such as storms and flooding, the rise of sea-levels or the immense losses of biodiversity, which have repercussions not only for the physical integrity of the planet but which pose direct threats to human life.\footnote{See for example \citet{gardiner2004} for a concise overview of the relevant climate science explained for non-climate scientists, or the IPCC and World Bank reports for more detail \citep{ipcc1.5spm,world2014turn}.}

Naturally, the public debate around this issue frequently invokes the question of \emph{responsibility}: Who carries how much backward-looking responsibility for the changes already inevitable, who is to blame; and who carries how much forward-looking responsibility to realise changes, who has to act?\footnote{What we call ``forward-looking'' or ex-ante responsibility is closely linked to the idea of obligation or duty, whereas what we call ``backward-looking'' or ex-post responsibility has also been called accountability, and relates to blame \citep{brahamvanhees2018,tamminga2019irreducibility}.}
And to what degree? In order to answer these questions, we need to clarify what exactly is meant by the term \emph{responsibility}, to consequently work on devising a proper quantification. Taxonomies of responsibility distinguish between six, nine, or even fifteen different concepts all denominated by the same term \citep{fischertognazzini2011,vandepoel2011,vincent2011}. We need to specify which specific meaning is being referred to before proceeding to elaborate on a quantification, both of which we aim to achieve through an appropriate formalisation.


\subsection{Existing work} 
The existing body of work regarding the question of how to correctly model responsibility in the context of climate change can roughly be divided into two categories, via the perspective from which this question is addressed. On the one side there are considerations focusing on applicability in the climate change context, computing tangible responsibility scores for countries or federations, with the aim of shaping the actions being taken and a lesser focus on conceptual elegance and generalisability \citep{botzen2008,mueller2009}. On the other side there is considerable work in formal ethics, aiming at understanding and formally representing the concept of responsibility in general with a special focus on rigour and well-foundedness, making it harder to account for messy real world scenarios (in realistic computation time) \citep{stit2001,brahamvanhees2012,chocklerhalpern2004,horty2001}.

It will be useful to highlight certain aspects of these works now. In the former set of works, and particularly also in public discourse, the degree of backward responsibility for climate change of a person, firm, or country is simply equated to cumulative past GHG emissions, or a slight variation of this measure \citep{guardian2011responsibility}. Certainly, this approach has one clear benefit, namely that it is easy to compute on any scale, and also extremely easy to communicate to a non-scientific audience. 
Similarly, certain authors assume a country's degree of forward responsibility to be proportional to population share, gross domestic product or some similar indicator, 
specifically in the debate about ``fair'' emissions allowances or caps \citep{poetter2019,ringius2002burden}.
However, unfortunately, such ad hoc measures clearly fail to generalise to other fields of application.

In the latter body of work, a principled approach is taken. Starting from considerations regarding the general nature of the concept of responsibility, formalisms are set up to represent these. These comprise causal models \citep{chocklerhalpern2004}, game-theoretical representations \citep{brahamvanhees2012,yazdanpanahdastani2016} or logics \citep{broersen2011mensrea,tamminga2019irreducibility}. A vast number of different aspects have been included in certain formalisations, such as degrees of causation or responsibility, relations between individuals and groups, or epistemic states of the agents to name but a few. Generally, these are discussed using reduced, well-defined example scenarios and thought experiments capturing certain complicating aspects of responsibility ascription. 

Additionally, there are investigations into the everyday understanding of the various meanings of the term `responsibility' \citep[e.g.]{vincent2011} as well as empirical studies regarding agents' responsibility judgements in certain scenarios, showing a number of asymmetry results \citep[e.g.]{nelkin2007}. However, we are here not concerned with mirroring agent's actual judgements, but rather with a normative account, so we will not go into detail about these.

\subsection{Research question}
As was stated above, we are interested in a formal representation and quantification of the concept of both forward- and backward-looking responsibility, with a specific focus on the application in the climate change scenario. That is, we aim to bridge between the two lines of approaches stated above, by staying within in the category of a principled and formal approach, but keeping in mind the practical applicability in complex scenarios.
Also, we want to relocate the space of discussion in the formal community by proposing a set of responsibility functions that differs from existing accounts in two important ways. 
One criterion which is generally agreed to be a precondition for responsibility ascription is some sort of relevant causal connection between the agent and the outcome, with the specific interpretations of the terms `relevant' and `causal' being up for discussion. In contrast to the existing formal accounts of responsibility, where causality is treated via a combination of necessity and sufficiency of sets of events for the occurrence of the outcome, we instead take a probabilistic approach. As a consequence, we are able to directly compute comparable numbers rather than having to manually extend a binary measure.
It might be useful to add that our work is normative, not descriptive. We aim at representing ways in which responsibility {\em should be} ascribed, not the ways in which people in standard discussion generally {\em do} ascribe it or are psychologically inclined to perceive it.  

We introduce a suitable framework that is able to represent all relevant aspects of a decision scenario using an extension of the game-theoretic framework of an extensive-form game tree. Subsequently, we will suggest candidate functions for assigning real numbers as degrees of responsibility (forward- as well as backward-looking) that are capable of correctly capturing the effects of previously discussed emblematic example scenarios and that will consequently be applied to a reduced representation of a climate change decision situation.





\subsection{Specific aspects to be considered} 
The above outline already shows several features of anthropogenic climate change that complicate responsibility assignments and occur in similar forms in other real-world decision problems.
We will now highlight and discuss several features that our framework will need to include, as well as certain aspects that we treat differently from existing work. 

\paragraph{Interaction, uncertainty.}
First of all, the effects of climate change are the result of an \emph{interaction} of many different actors: corporations, politicians, consumers, organisations, groups of these, etc.\ all play a role. Next, there is considerable \emph{uncertainty} regarding the impacts to be expected from a given amount of emissions or a given degree of global warming. While for some results we can assign probabilities and confidence intervals, for others this cannot be done in a well-founded way and beyond specifying the set of possible alternatives one cannot resolve the \emph{ambiguity} with the given state of scientific knowledge.

When several models give similar but slightly diverging predictions, for example, as is very often the case, we cannot assign probabilities to either of the models being `more right' than the others. What we can say however, is that each of the predictions is within the set of possible outcomes (given the premises, such as a certain future behaviour). The same goes for varying parameters within one and the same model.

Contrastingly, in a large body of work concerning effects of pollution, or warming, predictions are associated with a specified probability. Take for example the IPCC reports, such as the well known statement about the remaining carbon budget if warming is to be limited to 1.5 degrees: ``[\ldots] gives an estimate of the remaining carbon budget of [\ldots] 420 GtCO$_2$ for a 66\% probability [of limiting warming to 1.5\degree C]'' \citep{ipcc1.5spm}. 
In many cases, both aspects of uncertainty --- ambiguity and probabilistic uncertainty --- are combined by speaking about intervals of probabilities, which in particular the IPCC does pervasively \citep{ipcc_uncertain}.

We argue that it is equally important to take note of the additional information in the probabilistic uncertainty case\footnote{This is often called `risk' in economics, however, since we use the term `risky' in this article for a different concept, we stick to the term `probabilistic uncertainty' in order to ease disambiguation.} as of the lack thereof in the ambiguity case. It is known that the distinction between probabilistic and non-probabilistic uncertainty is important in decision making processes \citep{ellsberg1961}, and we want this to be reflected in our attribution of responsibility.

\paragraph{Nonlinearity.}
As a further particularity, the effects of global warming do not scale in a linear way with respect to emissions. With rising temperatures, so called `tipping elements' such as the Greenland or West Antarctic ice sheets risk being tipped \citep{lenton2019tippingpoints}: once a particular (but imprecisely known) temperature threshold is crossed, positive feedback leads to an irreversible progress of higher local temperatures and accelerated degradation of the element.\footnote{Ice reflects more sunlight than water. Thus, if a body of ice melts and turns into water, this will retain more heat than the ice did, leading to higher temperatures and faster melting of the remaining ice. As is stated in \citep{schellnhuberrahmstorf2016}: ``The keywords in this context are non-linearity and irreversibility''. Note that not all tipping elements are bodies of ice --- coral reefs or the Amazon rain forest also rank among them. The examples with corresponding explanation were chosen for their simplicity.} This initially local effect then aggravates global warming and may contribute to the tipping of further elements \citep{kriegler2009imprecise,wunderlingetal2019} adding up to the already immense direct impacts such as in case of these examples a sea level rise of several meters over the next centuries \citep{schellnhuberrahmstorf2016}.

We think that this nonlinearity should be reflected at least to some extent in the resulting responsibility attribution. This constitutes another argument for deviating from the --- linear --- cumulative past emissions accounts mentioned above \citep{botzen2008,mueller2009}.

\paragraph{Temporal progression.}
In contrast to existing formalisations of moral responsibility in game-theoretic terminology \citep{brahamvanhees2018}, we include a temporal dimension in our representation of multi-agent decisions by making use of extensive-form game trees rather than normal-form games. This temporal component is also featured in formalisations using the branching-time frames of {\em stit}-logics.\footnote{Note that normal-form game-theoretical models correspond to a subclass of \emph{stit} models \citep{duijf2018}. Similarly, extensive-form games can also be represented as a \emph{stit}-logic \citep{broersen2009}, but we don't pursue this further here as the additional features that we will include would complicate a logical representation and this is not currently necessary to express what we want to.}

\paragraph{Discounting of the future.}
However, we do not take into account the temporal distance of an outcome to the individual decisions that led to it. Unlike in the ongoing debate in the environmental economics community regarding the discounting factors to be employed when considering future damages, with the prominent opposition between William Nordhaus and Nicholas Stern \citep{nordhaus2007b,stern2010} and its ``non-decision'' by a large expert panel led by Ken Arrow \citep{arrow2013should}, our account is not directly affected by any form of discounting. 
This is due to the fact that while quantitative measures of welfare depend on notions of preference (which is arguably skewed towards the present), degrees of responsibility depend on causation (which does not decrease simply due to time\footnote{The reason that events that are very far away from an outcome, such as for example someone's being born versus their committing a crime later on in their life, tend to not be seen as causes is not due to the temporal distance, but to the interfering effects. As a parallel example, consider someone planting a bomb with a 50 year timer on it. This shows that it is not the temporal distance in itself that diminishes the causal connection, such as it is the case with preferences.}).
Still, if the effects of an action do disappear over time because of the underlying system dynamics (e.g., because pollutants eventually decay) and if this reduces the probability of causing harm much later, this fact can be reflected in the decision tree via probability nodes and the choice of the tree used to represent the situation in question.

\paragraph{Beliefs about other agents' actions.}
As another difference to existing formalisations we do not generally allow for assumptions regarding the likelihood of another agent's actions. The reason for this is that arguing with these probabilities enables evading a justified responsibility ascription. This can be seen for example in the shooting squad scenario, where an innocent prisoner is ordered to be shot by a group of soldiers, and where it is arguably very likely that at least one of the other soldiers shoots, but each individual shooter's responsibility should not be negated. Thus, while beliefs about others' actions may influence the psychologically perceived degrees of responsibility of the agents, it should not influence a normative assessment of their responsibility by an ideal ethical observer.

Note that even \citet{brahamvanhees2012}, who do consider beliefs about other's actions in their model, state that the important feature in a tragedy of the commons application is that ``[b]efore the game was played, each agent assigned at least some positive probability to the strategy combination the others actually did play''. That is, an unstructured set of possible outcomes suffices.

\paragraph{Causation.}
As mentioned previously, our account of causation differs from those used in many previous formal representations. Namely, while in most previous accounts the clear insufficiency of a but-for condition is avoided by employing NESS \citep{wright1988}  
or INUS \citep{mackie1965} conditions, these remain binary, deterministic and based on the concepts of necessity and sufficiency.
We instead use a probabilistic account, where an event is seen as a cause of an outcome to the degree that it raised its probability, as suggested for example by Vallentyne \citep{vallentyne2008bruteluck}.

The but-for condition claims that an event is a cause of an effect if the effect had not occurred, were it not for the event in question. This is at the same time too strict, failing to capture overdetermination, and too inclusive, including for example very indirect but necessary events such as someone's being born related to an action they perform in their 50's.  
NESS is an acronym for `necessary element of a sufficient subset', that is, an event is seen as a cause for an outcome if and only if it is a necessary element of a subset of all the events leading to the outcome, and the subset together is sufficient for the outcome to occur. For example, consider five people pushing a car down a cliff, but it would only take three people to move it. Then one single person pushing is a necessary element of the sufficient subset of any group of three people (that they are part of) pushing the car. 
Similarly, INUS is an acronym for `insufficient but necessary element of an unnecessary but sufficient set of events'. This view captures the intuition that an outcome might have several sets of events that can explain it, of which either is sufficient (but neither is necessary), and a cause is any necessary element of each of these sets\footnote{Reducing this notion to one possible set of explanatory events by removing the `insufficient' and `unnecessary' conditions results in the NESS condition}.

Despite discussions and suggestions for using probabilistic approaches to represent causation \citep{fenton-glynn2016,hitchcock2021,kvart2002}, these views have to our knowledge not yet been employed in generalised formalisations of responsibility.
We employ here the account as stated by \citet{vallentyne2008bruteluck} in relation to responsibility: ``I shall assume that the relevant causal connection is that the choice increases the objective chance that the outcome will occur --- where objective chances are understood as objective probabilities''. This account lends itself to our approach as we specifically want to discuss situations in which the outcome occurs with a given probability, and it enables a straightforward representation of {\em degrees} of responsibility. Note however, that unlike \citet{vallentyne2008bruteluck} we do not refer to agents' {\em beliefs} regarding the probabilities.

 
\subsection{Structure of the paper} 
The remainder of the paper is structured as follows. We will begin in Sect.~\ref{sec:paradigmatic} by discussing a set of paradigmatic example scenarios and the results that a responsibility function should assign in them.
In Sect.~\ref{sec:formalmodel} we continue  with a presentation of the framework we use to represent relevant decision scenarios in order to formulate responsibility functions. In Sect.~\ref{sec:rfunctions} we introduce four different candidate responsibility functions (all differentiated between backward- and forward-looking formulations) and evaluate them by determining to what extent they are able to correctly capture the aforementioned paradigmatic example scenarios. In Sect.~\ref{sec:application} we present a reduced real-world application scenario in which we apply the best-performing responsibility functions. Finally, we conclude in Sect.~\ref{sec:conclusion}.


\section{Paradigmatic examples} \label{sec:paradigmatic}


In this section we will have a look at a few paradigmatic example scenarios, mostly known from the literature, together with their proposed evaluation. We will then use these examples to assess our proposed formal representation by testing whether our model outputs the same results as the `analytic' evaluation we present here.

\subsection{Common scenarios and their evaluation}

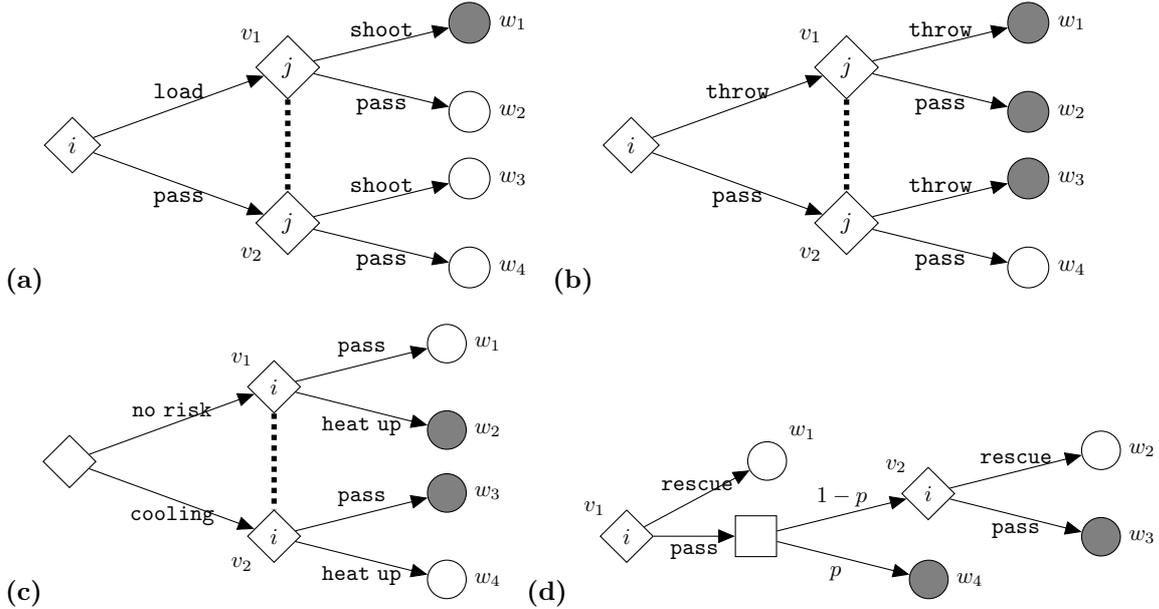
\begin{figure}
\subfigure{
    {\bf (a)}\begin{tikzpicture} [-, scale=.9, every node/.style={scale=.9}, node distance=1cm, decision node/.style={diamond, draw, aspect=1, minimum height= 7mm},  prob node/.style={regular polygon,regular polygon sides=4, draw, minimum height= 1cm},  dummy/.style = {}, good outcome/.style={circle, draw, minimum width=6mm}, bad outcome/.style={circle, draw, fill=gray, minimum width=6mm }, >= triangle 45]
\node[decision node] (1) [] [] {$i$};

\node[dummy] (a) [right of = 1] [] {};
\node[dummy] (a2) [right of = a] [] {};

\node[decision node] (2) [above right =  of a2] [label=above left: $v_1$] {$j$};
\node[decision node] (3) [below right = of a2] [label=below left: $v_2$] {$j$};

\node[dummy] (b) [right of = 2] [] {};
\node[dummy] (b2) [right of = b] [] {};
\node[dummy] (b3) [right of = b2] [] {};

\node[bad outcome] (good2) [above right = .4cm of b2] [label = right: $w_1$]{};
\node[good outcome] (bad2) [below right = .4cm of b2] [label = right: $w_2$] {};

\node[dummy] (c) [right of = 3] [] {};
\node[dummy] (c2) [right of = c] [] {};

\node[good outcome] (bad3) [above right = .4cm of c2] [label = right: $w_3$] {};
\node[good outcome] (good3) [below right = .4cm of c2] [label = right: $w_4$] {};

\path[->] (1) edge node[above] {$\mathtt{load}$}  (2) (1) edge node[below] {$\mathtt{pass}$} (3) (2) edge node[above] {$\mathtt{shoot}$}  (good2) (2) edge node[below] {$\mathtt{pass}$} (bad2) (3) edge node[below] {$\mathtt{pass}$} (good3) (3) edge node[above] {$\mathtt{shoot}$} (bad3);

\draw[line width=2pt, dotted=on] (2) -- (3); 

\end{tikzpicture}}
\subfigure{
    {\bf (b)}
    \begin{tikzpicture} [-, scale=.9, every node/.style={scale=.9}, node distance=1cm, decision node/.style={diamond, draw, aspect=1, minimum height= 7mm},  prob node/.style={regular polygon,regular polygon sides=4, draw, minimum height= 1cm},  dummy/.style = {}, good outcome/.style={circle, draw, minimum width=6mm}, bad outcome/.style={circle, draw, fill=gray, minimum width=6mm }, >= triangle 45]
\node[decision node] (1) [] [] {$i$};

\node[dummy] (a) [right of = 1] [] {};
\node[dummy] (a2) [right of = a] [] {};

\node[decision node] (2) [above right =  of a2] [label=above left: $v_1$] {$j$};
\node[decision node] (3) [below right = of a2] [label=below left: $v_2$] {$j$};

\node[dummy] (b) [right of = 2] [] {};
\node[dummy] (b2) [right of = b] [] {};
\node[dummy] (b3) [right of = b2] [] {};

\node[bad outcome] (good2) [above right = .4cm of b2] [label = right: $w_1$]{};
\node[bad outcome] (bad2) [below right = .4cm of b2] [label = right: $w_2$] {};

\node[dummy] (c) [right of = 3] [] {};
\node[dummy] (c2) [right of = c] [] {};

\node[bad outcome] (bad3) [above right = .4cm of c2] [label = right: $w_3$] {};
\node[good outcome] (good3) [below right = .4cm of c2] [label = right: $w_4$] {};

\path[->] (1) edge node[above] {$\mathtt{throw}$}  (2) (1) edge node[below] {$\mathtt{pass}$} (3) (2) edge node[above] {$\mathtt{throw}$}  (good2) (2) edge node[below] {$\mathtt{pass}$} (bad2) (3) edge node[below] {$\mathtt{pass}$} (good3) (3) edge node[above] {$\mathtt{throw}$} (bad3);

\draw[line width=2pt, dotted=on] (2) -- (3); 
\end{tikzpicture}}

\subfigure{
    {\bf (c)}\begin{tikzpicture} [-, scale=.85, every node/.style={scale=.85}, node distance=1cm, decision node/.style={diamond, draw, aspect=1, minimum height= 7mm},  prob node/.style={regular polygon,regular polygon sides=4, draw, minimum height= .9cm},  dummy/.style = {}, good outcome/.style={circle, draw, minimum width=6mm}, bad outcome/.style={circle, draw, fill=gray, minimum width=6mm }, >= triangle 45]
\node[decision node] (1) [] [] {$\ \ \ $};

\node[dummy] (a) [right of = 1] [] {};
\node[dummy] (a2) [right of = a] [] {};

\node[decision node] (2) [above right =  of a2] [label=above left: $v_1$] {$i$};
\node[decision node] (3) [below right = of a2] [label=below left: $v_2$] {$i$};

\node[dummy] (b) [right of = 2] [] {};
\node[dummy] (b2) [right of = b] [] {};
\node[dummy] (b3) [right of = b2] [] {};

\node[good outcome] (good2) [above right = .4cm of b2] [label = right: $w_1$]{};
\node[bad outcome] (bad2) [below right = .4cm of b2] [label = right: $w_2$] {};

\node[dummy] (c) [right of = 3] [] {};
\node[dummy] (c2) [right of = c] [] {};

\node[bad outcome] (bad3) [above right = .4cm of c2] [label = right: $w_3$] {};
\node[good outcome] (good3) [below right = .4cm of c2] [label = right: $w_4$] {};

\path[->] (1) edge node[above] {$\mathtt{no\ risk} $}  (2) (1) edge node[below] {$\mathtt{cooling}$} (3) (2) edge node[above] {$\mathtt{pass}$}  (good2) (2) edge node[below] {$\mathtt{heat\  up} $} (bad2) (3) edge node[below] {$\mathtt{heat\ up} $} (good3) (3) edge node[above] {$\mathtt{pass} $} (bad3);

\draw[line width=2pt, dotted=on] (2) -- (3); 

\end{tikzpicture}}
\subfigure{
    {\bf (d)}
    \begin{tikzpicture} [-, scale=.85, every node/.style={scale=.85}, node distance=1cm, decision node/.style={diamond, draw, aspect=1, minimum height= 7mm},  prob node/.style={regular polygon,regular polygon sides=4, draw, minimum height= .9cm},  dummy/.style = {}, good outcome/.style={circle, draw, minimum width=6mm}, bad outcome/.style={circle, draw, fill=gray, minimum width=6mm }, >= triangle 45]
\node[decision node] (1) [] [label=above left: $v_1$] {$i$};

\node[dummy] (a) [right of = 1] [] {};

\node[good outcome] (2) [above right =  of a] [label=above right: $w_1$] {};
\node[prob node] (3) [right of = a] [] {};

\node[dummy] (b) [right of = 3] [] {};
\node[dummy] (b2) [right of = b] [] {};

\node[decision node] (4) [above right = .4cm of b2] [label=above left: $v_2$]{$i$};
\node[bad outcome] (bad) [below right = .4cm of b2] [label = right: $w_4$] {};

\node[dummy] (c) [right of = 4] [] {};
\node[dummy] (c2) [right of = c] [] {};

\node[good outcome] (good2) [above right = .4cm of c2] [label = right: $w_2$] {};
\node[bad outcome] (bad2) [below right = .4cm of c2] [label = right: $w_3$] {};

\path[->] (1) edge node[above] {$\mathtt{rescue}$}  (2) (1) edge node[below] {$\mathtt{pass}$} (3) (3) edge node[above] {$1-p$} (4) (3) edge node[below] {$p\ \ $} (bad) (4) edge node[above] {$\mathtt{rescue}$} (good2) (4) edge node[below] {$\mathtt{pass}$} (bad2) ;

\end{tikzpicture}}
\caption{\label{fig:paradigmatic} Paradigmatic example scenarios. Diamond-shaped nodes labelled with an agent represent a decision node of this agent, with outgoing arrows labelled with the available actions. Empty diamonds represent ambiguity nodes. Empty squares represent probabilistic uncertainty nodes, with the outgoing edges labelled with the corresponding probability. Circles represent outcomes, with those shaded in being undesirable. Dotted lines between nodes represent indistinguishability for the deciding agent. 
(a) Load and shoot. One agent $i$ may load a gun or not before passing it to another agent $j$ who can shoot an innocent victim or not.
(b) Rock throwing. Two agents have the choice to throw a rock into a window, thus shattering it. Agent $j$ does not know whether the other agent, $i$, already threw their stone.
(c) Cooling vs. Heating. Agent $i$ can select whether to induce global warming, not knowing whether there is a threat of an impending ice age that this would alleviate or not, in which case the warming leads to a climate crisis.
(d) Hesitation. Agent $i$ can choose to save someone from drowning immediately, or to hesitate. If they hesitate they might get another chance to save the person from drowning with a certain probability, but it might also be too late.}
\end{figure}

The `training set' examples have been selected because they each represent an interesting aspect of responsibility attribution that we will want to capture correctly with the proposed responsibility functions. Additionally, some scenarios have been adjusted in order to showcase every possible kind of uncertainty node that we make available in our framework. Let us now describe the scenarios. Graphical representations can be seen in Fig.~\ref{fig:paradigmatic}.

\begin{itemize}
    \item {\bf Load and shoot.} Agent $i$ has the choice to either load a gun that they pass to a second agent, $j$, or to not load it. Agent $j$ then has the choice to shoot an innocent victim with said gun or not, not knowing whether the gun is loaded. Represented in Fig.~\ref{fig:paradigmatic}(a).
    \item {\bf Rock throwing.} Agent $i$ has the choice to either throw a rock into a window, thus shattering the window, or not. A second agent $j$ then has the same choice, not knowing whether agent $i$ already threw a stone before them. Represented in Fig.~\ref{fig:paradigmatic}(b).
    \item{\bf Cooling vs. Heating.} It is unclear whether humanity (agent $i$) is faced with an impending ice age or not. They nevertheless have to select whether or not to heat up the earth through green house gas emissions, which, in case of an absent ice age, will lead to catastrophic heating. Represented in Fig.~\ref{fig:paradigmatic}(c).
    \item {\bf Hesitation.} Agent $i$ has the choice to either rescue an innocent stranger immediately, or hesitate, in which case they might get another chance at rescuing the innocent stranger at a later stage, but it might also already be too late.\footnote{While this example might seem somewhat odd in direct interaction contexts --- imagine the scenario described above where someone has the choice to save a person from drowning immediately or first finish off their ice-cream knowing that with probability $p$ the other person will drown in the meantime --- it represents a common issue in climate change mitigation efforts.} Represented in Fig.~\ref{fig:paradigmatic}(d).
\end{itemize}

We argue that these examples should be analysed as follows. These analyses are closely related to questions of moral luck \citep{Andre1983,Nagel1979,Tong2004},
reasonable beliefs \citep{baron2016justification}, and ignorance as an excuse \citep{zimmerman2016ignorance}.

\paragraph{Load and shoot.}

The examples {\em Load and shoot} and {\em Rock throwing} are parallel to one another, both including situations in which agent $j$ might not actually be able to influence the outcome (because either the gun is not loaded so it does not matter whether they shoot or not, or because the other agent already threw a stone that will shatter the window), but they do not know whether they are in this situation or in the one where their action does have an impact. In both cases we argue that the responsibility ascription must take into account the viable option that the agent's action will have/would have had an impact. Therefore, the agent cannot dodge responsibility by referring to this uncertainty. They should be assigned full forward and backward (if they select the possibly harmful action) responsibility. This relates to the discussion about moral luck, and the case for disregarding factors that lie outside of the agent's control is argued in \citet{Nagel1979}: ``Where a significant aspect of what someone does depends on factors beyond [their] control, yet we continue to treat [them] in that respect as an object of moral judgment, it can be called moral luck. Such luck can be good or bad. [\dots] If the condition of control is consistently applied, it threatens to erode most of the moral assessments we find it natural to make.''

This also relates to a prominent criticism of the probability raising account for causation, namely that an agent may raise the probability of an event without this event actually occurring as a result, as the probability stayed below 1. Similarly to situations in which the event does not end up occurring due to the actions of others that the agent had no knowledge of or influence over, we argue that this should not reduce responsibility ascription but rather lead to what can be interpreted as a form of `counterfactual' responsibility.

\paragraph{Rock throwing.} As stated previously, this example is symmetric to the `load and shoot' one. Again, agent $j$ does not know whether they are in a situation in which their action can have an effect on the outcome or not. We argue that due to this lack of knowledge, agent $j$ must assume the possibility that their actions will have negative consequences. Thus, again, they should be assigned full forward and backward (if they choose to throw the rock) responsibility. This analysis also relates to the issue of assigning probabilities to other's actions. If $j$ were able to plausibly assume that $i$ did/did not throw the rock with probability 50/50 or 90/10, then this could be taken into account for a computation of partial responsibility instead of staying within the absolute case as we do. However, recall that we exclude such a reasoning, as argued in Sect.~\ref{sec:intro}. 

\paragraph{Cooling vs. Heating.} This situation is somewhat non-trivial with respect to responsibility ascription. In the two cases, if there is an impending ice age and if there is none, the agent has the choice between the same actions, but the actions have directly opposing outcomes. Additionally, the agent does not know which situation they are in. Therefore, no matter what they do, they must always assume that their action may lead to the undesired outcome, but they also must always assume that it might be the only option to achieve the desired outcome. Whether or not we should assign  responsibility in all nodes, or in none of them, depends on whether we want to prevent responsibility voids, or whether we want to ensure agents always have the option to avoid  responsibility.

\paragraph{Hesitation.} This scenario helps determine the effect of repeated actions and shows whether it is possible to redeem a negatively evaluated action. Let us first compare the backwards-looking responsibility in the two morally desirable outcomes: $w_1$, where the agent decided to immediately save the stranger, and $w_2$, where the agent first hesitated but was lucky to get a second chance and ended up rescuing the stranger. Clearly, if the agent decides to immediately rescue the stranger, they should be assigned zero backwards-looking responsibility. However, as we discussed previously, events which the agent has no influence over, such as the outcome of a probabilistic event, should not modify responsibility ascription. Therefore, even if the agent does in the end save the stranger after an initial hesitation, we argue that they should be assigned at least some backwards-looking responsibility for risking the stranger's death. This differs from other accounts that put more emphasis on the actualised outcome, such as  \citet{braham2009degrees}.\footnote{Note that according some accounts, such as the one applied in the parable of the lost son in the New Testament, one might even consider that the actions leading to $w_2$ are morally superior to those leading to $w_1$. That is, this account claims it is more praiseworthy to select a morally reproachable action and subsequently select that action which still results in the good outcome, rather than to have immediately selected the morally praiseworthy action.} Similarly, if we look at the undesirable outcome $w_4$ where the agent also initially hesitated but was not lucky and the stranger died, we argue that they should not carry full responsibility for the stranger's death, as there was the possibility they might get another chance at rescuing them. Lastly, in the situation where the agent repeatedly decided not to help, they should be assigned full responsibility. This is again quite clear. With respect to forward-looking responsibility, the situation again requires some thought. The second decision node does not differ, in a forward-looking way, from any other node which offers the decision between an action that immediately leads to a desired outcome and one that immediately leads to an undesired outcome. Therefore, in this node the agent has full forward-looking responsibility. Contrastingly, in the first node the agent has the choice between an action that immediately leads to a good outcome and a probabilistically uncertain outcome. While it might seem like the agent does not have full influence over the situation, we argue that indeed they do, as the second decision node belongs to the same agent, and agent $i$ can therefore ensure both that the stranger survives, by immediately rescuing them, and that they die, by passing twice. Thus, we argue that agent $i$ carries full forward-looking responsibility in this case.

\section{Formal model}\label{sec:formalmodel}

In this section we will propose a formal framework for the study of responsibility in multi-agent settings with various forms of uncertainty. For reasons of completeness, we spell out the full formal representation. However, readers unfamiliar with this notation may wish to skip those parts and instead focus on the interpretation and the graphical representation.

In order to represent decisions of an interacting set of agents in a temporal progression, the game-theoretical data structure of a game in extensive form has proven effective. We therefore base our representation on this structure. However, we need to extend it in order to be able to represent all the features described above. Specifically, we add special types of nodes for encoding uncertainty.
Also, in contrast to games, our structures do not specify individual payoffs for all outcomes
but only a set of ethically undesired outcomes.
This is sufficient, as we will not apply any game-theoretic analyses referring to rational courses of actions or utility maximisation but rather use this data structure to talk about responsibility assignments.

\subsection{Framework}

We use $\Delta(A)$ to denote the set of all probability distributions on a set $A$.

\paragraph{Trees.}
We define a {\em multi-agent decision-tree with ambiguity} (or shortly, a {\em tree}) to be a structure\\ 
$\T = \langle I,(V_i),V_a,V_p,V_o,E,\sim,(A_v),(c_v),(p_v) \rangle$ 
consisting of:
\begin{itemize}
\item A nonempty finite set $I$ of {\em agents} (or players).
\item For each agent $i\in I$, a finite set $V_i$ of $i$'s {\em decision nodes}. At each decision node, only one agent acts, so all these sets $V_i$ are disjoint.
      We denote the set of all decision nodes by
      $V_d \eqdef  \bigcup_{i\in I} V_i$.
\item Further disjoint finite sets of nodes:
      a set $V_a$ of {\em ambiguity nodes}, 
      a set $V_p$ of {\em probability nodes},
      and a nonempty set $V_o$ of {\em outcome nodes}.
      We denote the set of all nodes by $V \eqdef  V_d \cup V_a \cup V_p \cup V_o$.
\item A set of directed edges $E\subset V\times V$ so that $(V,E)$ is a directed tree 
      whose leaves are exactly the outcome nodes:
      $$ V_o = \{ v\in V : \not\exists v'\in V ((v,v')\in E) \}. $$
      For all $v\in V\setminus V_o$, let $S_v \eqdef  \{ v'\in V: (v,v')\in E \}$ denote the set of {\em possible successor nodes} of $v$.
\item An {\em information equivalence} relation $\sim$ on $V_d$ so that $v \in V_i $ and $v'\sim v$ implies $v'\in V_i$.
      We call the equivalence classes of $\sim$ in $V_i$ the {\em information sets} of $i$. 
\item For each agent $i\in I$ and decision node $v\in V_i$, 
      a nonempty finite set $A_v$ of $i$'s possible {\em actions} in $v$,
      so that $A_v = A_{v'}$ whenever $v\sim v'$, 
      and a bijective {\em consequence} function mapping actions to successor nodes, $c_v:A_v\to S_v$.
\item For each probability node $v\in V_p$, a probability distribution $p_v\in\Delta(S_v)$ 
      on the set of possible successor nodes. 
\end{itemize}
Our interpretation of these ingredients is the following:
\begin{itemize}
\item A tree encodes a multi-agent decision situation where agents can make choices in a given order.
      An outcome node $v\in V_o$ represents an ethically relevant state of affairs 
      that may result from these choices.
\item Each decision node $v\in V_i$ represents a point in time where agent $i$ has the agency to make a decision at free will.
      The elements of $A_v$ are the mutually exclusive choices $i$ can make, including any form of ``doing nothing'',
      and $c_v(a)$ encodes the immediate consequences of choosing $a$ in $v$. 
      Often, $c_v(a)$ will be an ambiguity or probability node to encode uncertain consequences of actions.
\item Probability and ambiguity nodes as well as information equivalence are used to represent various types of uncertainty. 
      The agents are assumed to always commonly know the full tree.
      For any probability node $v \in V_p$, they know that
      the possible successor nodes are given by $S_v$ and have probabilities $p_v(v')$ for $v'\in S_v$.       
      Note that these probabilities are universal, rather than individual.
      About an ambiguity node $v\in V_a$ they only know that
      the possible successor nodes are given by $S_v$, without being able to rightfully attach probabilities to them. 
      Ambiguity nodes can also be thought of as decision nodes associated to a special agent called `nature'.
      
      In contrast to the universal uncertainty at the tree-level encoded by probability and ambiguity nodes, information-equivalence is used to encode uncertainty at the agent level.       
      An agent $i$ cannot distinguish between their information-equivalent decision nodes $v\sim v'$, with $v,v'\in V_i$. Information-equivalent nodes $v \sim v'$ have the same set of possible actions $A_v = A_v'$, as otherwise the agent would be able to distinguish between them via the available actions.
\end{itemize}
An ambiguity node whose successors are probability nodes can be used to encode uncertain probabilities
like those reported by the IPCC \citep{ipcc_uncertain} 
or those corresponding to the assumption that ``nature'' uses an Ellsberg strategy \citep{Decerf2019}.

Note that in contrast to some other frameworks, e.g., those using {\em normal}-form (instead of extensive-form) game forms such as \citep{brahamvanhees2018},
our trees do not directly allow for two agents to act at the exact same time point.
Indeed, in a real world in which time is continuous, one action will almost certainly precede another, if only by a minimal time interval. 
Still, 
two actions may be considered ``simultaneous'' 
if they occur so close in time that the later acting player cannot know what the earlier action was. This ignorance can easily be encoded by means of information equivalence in a way similar to Fig.\ \ref{fig:paradigmatic} (a) -- (c).

As with any modelling exercise, the result of the representation depends strongly on the point of view of the modeller and the corresponding selection of what to include in the representation.
If the modeler follows the basic idea that what matters for the assessment of an agent $i$'s responsibility is what $i$ ``reasonably believes'' in any decision node $v_d$ (as in \citet{baron2016justification}),
then the available actions, consequences thereof and probability distributions should be represented accordingly.
If, on the other hand, the modeler follows the idea that some ``objective reality'' is the basis for evaluation, even if it might have been unknown to the agent in question (as for the strict liability described by \citet{broersen2011mensrea} or the causal responsibility concept mentioned in \citet{vallentyne2011}) then this, too, will be reflected in the available actions and consequences included in the model. 

\paragraph{Events, groups, responsibility functions.}
As in probability theory, we call each subset of outcomes a possible {\em event.}
In the remainder of this paper, we will use $\eps$, with $\eps \subseteq V_o$ to represent an ethically {\em undesirable} event, those events not in $\eps$ are considered ethically desirable.
Any nonempty set $G\subseteq I$ of agents is called a {\em group} in this article.\footnote{Note that we deliberately do not require that a set of agents shares any identity or possesses ways of communication or coordination for an ethical observer to meaningfully attribute responsibility to this ``group''.}

Our main objects of interest are quantitative metrics of degrees of responsibility that we formalise as 
{\em backward-looking responsibility functions} $\R_b$ 
and {\em forward-looking responsibility functions} $\R_f$. 

A backward-looking responsibility function maps a group $G$, an outcome node $v \in V_o$, the decision situation $\T$ and the undesirable outcome $\eps$ to a real number $\R_b(\T,v,G,\eps)$. 
It is meant to represent some form of degree of backward-looking responsibility of $G$ regarding $\eps$ 
in the multi-agent decision situation encoded by $\T$ when outcome $v$ has occurred.

A forward-looking responsibility function maps every combination of tree $\T$, group $G$, event $\eps$, and {\em decision} node $v\in V_d$ 
to a real number $\R_f(\T,v,G,\eps)$
meant to represent some form of degree of forward-looking responsibility of $G$ regarding $\eps$ 
in the multi-agent decision situation encoded by $\T$ when in decision node $v$.

When we are concerned with only one agent, i.e. $G=\{i\}$, we also write $\R_{b/f}(\T,v,i,\eps)$.
Whenever any of the arguments $\T$, $v$, $G$, $\eps$ are kept fixed and are thus obvious from the context,
we omit to explicate them when writing $\R_{b/f}$ or any of the auxiliary functions defined below.

\paragraph{Graphical representation.}
As exemplified in Fig.\ \ref{fig:paradigmatic}, we can represent a tree $\T$ and event $\eps$ graphically as follows.
Edges are shown as arrows, 
decision nodes are diamonds labelled by agents, with outgoing arrows labelled by actions, 
ambiguity nodes are unlabelled diamonds,
probability nodes are squares with outgoing arrows labelled by probabilities,
and outcome nodes are circles, filled in grey if the outcome belongs to $\eps$.
Finally, information equivalence is indicated by dashed lines connecting or surrounding the equivalent nodes. 

\paragraph{Auxiliary notation.}
The set of decision nodes of a group $G\subseteq I$ is the union of the decision nodes of its members, $V_G \eqdef  \bigcup_{i\in G}V_i$.
To ease the definition of ``scenario'' below we denote the set of nodes that are not in $V_G$ and are not probabilistic uncertainty nodes (i.e., all non-$G$ decision nodes and all ambiguity nodes) by $V_{-G} \eqdef  V_d \setminus ( V_G \cup V_a)$. This can be thought of as nodes where someone who is not part of group G -- another agent or Nature -- makes a decision.

The {\em history} of a node $v\in V$ are all those nodes that lie on the path from the start node to this node. If $v'\in S_v$, we call $P(v') \eqdef v$ the {\em predecessor} of $v'$.
Let $v_0 \in V$ be the {\em root} node of $(V,E)$, i.e., the only node without predecessor.
The {\em history} of $v\in V$ is then $H(v) \eqdef  \{ v, P(v), P(P((v)), \dots, v_0 \}$.
In the other direction, we call the nodes accessible after $v$ $B(v) \eqdef  \{v'\in V: v\in H(v')\}$ the {\em (forward) branch} of $v$. Note that there is only one past but possibly a branching future.
Taking into account information equivalence, we also define the {\em information branch} of $v$ 
$B^\sim(v) \eqdef  \bigcup_{v'\sim v} B(v')$, as all those nodes that, for all the player deciding in $v$ knows, may be reachable from their current position. 

For a node $v\in V$, a decision node in its history $v'\in H(v)\cap V_d$ and that action $a$ available in $v'$ which takes us on the path to $v$, i.e. $c_{v'}(a)\in H(v)$, we introduce the notation $a_{v' \to v}$.

A decision node $v\in V_d$ where the agent has no information uncertainty $\{v':v'\sim v\}=\{v\}$ is called a {\em complete information} node.

From now on, unless stated otherwise, we will use $w$ to refer to outcome nodes $w\in V_o$ and $v$ to refer to non-outcome nodes $v\in V\setminus V_o$. Mostly, $v$ will be used to designate decision nodes $v\in V_d$.

\paragraph{Strategies, scenarios, likelihoods.}

In order to be able to quantify responsibility ascription, we will need to be able to range over all the ways that the uncertain evolution of the future plays out. To this end, we will use the notions of \emph{strategy} and \emph{scenario}, with a strategy telling us what agents within a given group will be doing, and a scenario resolving all other unquantified uncertainty by telling us what agents outside of the group (including `nature') will be doing as well as revealing the current node within an information set. We do not need to resolve quantified probabilistic uncertainty, as we can use \emph{likelihoods} in the computation of responsibility.

Formally, given a tree $\T$, a group $G\subseteq I$, and a node $v\in V$, a \emph{strategy} for $G$ at $v$ is a function \[ \sigma \colon V_G \cap B^{\sim}(v) \to \bigcup_{v' \in V_G\cap B^{\sim}(v)} A_{v'} \]
that chooses an action for every one of $G$'s future decision nodes considered possible according to the information set in $v$.  Let $\Sigma(\T,G,v)$ (or shortly $\Sigma(v)$ if $\T,G$ are fixed) be the set of all those strategies.
For $\sigma\in\Sigma(\T,G,v)$, let 
$$ V_o^\sigma \eqdef \{ w\in B^\sim(v)\cap V_o : a_{v'\to w} = \sigma(v')\text{~for all~}v'\in B^\sim(v)\cap H(w)\cap V_G \}, $$
i.e., the set of possible outcomes when group $G$ follows strategy $\sigma$ from node $v$ on.

Complementarily, a \emph{scenario} for $G$ at $v$ is a node $v_{\zeta} \sim v$ together with a function \[ \zeta \colon V_{-G} \cap B(v_{\zeta}) \to \bigcup_{v'\in V_{-G} \cap B(v_{\zeta})} S_{v'} \]
such that $\zeta(v')=c_{v'}(a)$ and $\zeta(v'')=c_{v''}(a)$ for some $a\in A_{v'}$ whenever $v'\sim v'' \in V_{-G} \cap B(v_{\zeta}) $.

The node $v_{\zeta}$ resolves uncertainty regarding the current information set, while the function $\zeta$ resolves non-probabilistic uncertainty regarding what will happen in the future (disregarding $G$'s own actions), by choosing successor nodes $\zeta (v')$ for all future decision nodes by non-members of G or ambiguity nodes. The condition on information-equivalent nodes states that a scenario cannot differentiate between them and select two different actions in each node but must select the same one.\\
Let $Z^\sim(\T,v,G)$ (or shortly $Z^\sim(v)$) be the set of all scenarios at $v$
and $Z(\T,v,G)\subseteq Z^\sim(\T,v,G)$ (or shortly $Z(v)$) that of all scenarios at $v$ with $v_\zeta=v$.

Given a starting node $v$, we can define a probability distribution $\pi_{v}\in\Delta(V_o\cap B^\sim(v))$ on the set of possible outcome nodes in the following straightforward way:
\begin{align}
    \psi(v) &= 1, \\
    \psi(v'') &= \psi(v')\quad \text{for~} v'\in V_d \text{ or } v'\in V_a,\ v'' \in S_{v'},\\
    \psi(v'') &= \psi(v') p_{v'}(v'')\quad \text{for~} v'\in V_p, v''\in S_{v'}, \\
    \pi_{v}(w) &= \psi(w) \quad \text{for all~} w\in V_o\cap B^\sim(v).
\end{align}

Let us denote the resulting likelihood of $\eps$ given the starting node $v$ by 
$$ \ell(\eps|v) \eqdef \sum_{w\in\eps}\pi_{v}(w). $$

Consequently, to define the likelihood of $\eps$ given a starting node $v$ together with scenario $\zeta$ and strategy $\sigma$, we will in each step consider only those successor nodes that can be reached when following $\sigma$ or $\zeta$, and assign zero likelihood to all other nodes:

\begin{align}
    \chi(v_{\zeta}) &= 1, \\
    \chi(v'') &= \chi(v')\quad \text{for~} [ v'\in V_G \text{ and } v''= c_{v'}(\sigma(v'))] \text{ or } [v'\in V_{-G}\text{ and } v'' = \zeta(v')],\\
    \chi(v'') &= \chi(v') p_{v'}(v'')\quad \text{for~} v'\in V_p,\  v''\in S_{v'}, \\
    \chi(v'') &= 0\quad \text{for all other~} v''\in B^\sim(v), \\
    \pi_{v, \sigma, \zeta}(v_o) &= \chi(w) \quad \text{for all~} w\in V_o\cap B^\sim(v).
\end{align}

Recall that $v_{\zeta}$ is that node in $v$'s information set which according to scenario $\zeta$ is the actual one. Only this node, and not $v$ itself, is relevant in the computation of $\chi$.

Again, we denote the resulting {\em likelihood of $\eps$ given starting node $v$, strategy $\sigma$ and scenario $\zeta$ }by 
 $$ \ell(\eps|v,\sigma,\zeta) \eqdef \sum_{w\in\eps}\pi_{v,\sigma,\zeta}(w). $$


\section{Candidate responsibility functions}\label{sec:rfunctions}

In this section we will introduce four pairs of responsibility functions $(\R_f, \R_b)$, that measure degrees of responsibility in terms of 
differences in likelihoods of the undesired outcome, together with a reference function $\R_b^0$ related to strict causation. 
We will examine whether the proposed functions output the right results for the paradigmatic example scenarios described and analysed in section~\ref{sec:paradigmatic}, and see that we need to progressively adjust them in order to achieve the desired outcomes.

In general, we will mostly view backwards responsibility as some measure related to `having influenced the outcome towards the undesired one', for example by increasing likelihood or risk-taking. Correspondingly, we view forward-looking responsibility as a measure related to influence. Thus, the forward responsibility in a certain node will often be the maximal possible value of backwards responsibility arising from actions taken in that node. 

To simplify the notation we will use some conventions in the remainder of this chapter.
Let $\T$, $G$, and $\eps$ be given; we will drop them from notation.
With respect to the various different nodes appearing in the definitions, we stick with the previously mentioned convention and denote the decision node at which $\R_f$ is evaluated by $v\in V_d$, the outcome node at which $\R_b$ is evaluated by $w\in V_o$, and other nodes also by $v\in V$ or $ v', v''\in V$ so that $v$ comes before $v'$ comes before $v''$ (i.e., $v\in H(v')$, $v'\in B(v)$ etc.).

\paragraph{Benchmark variant: strict causation.}

The most straightforward definition of 
a forward responsibility function in our framework that resembles the strict causation view, as employed for example in the most basic way of (being able to) `see to it that' is to set
\[ \R^0_f(v) \eqdef
\begin{cases}
1 \text{ iff } B(v)\cap V_o\not\subseteq\eps \text{ and } B(c_{v}(a))\cap V_o\subseteq\eps \text{ for some } a\in A_{v},\\
0 \text{ else}
\end{cases}
\]
that is, if at decision node $v_d$, the undesired outcome is not certain, but there is an action the agent can select which makes it certain.
The corresponding backward responsibility function would be to assign full responsibility to G if at any point they toggled the undesired outcome from uncertain to happening.
\[ \R^0_b(w) \eqdef
\begin{cases}
1 \text{ iff } B(v)\cap V_o\not\subseteq\eps \text{ and } B(v')\cap V_o\subseteq\eps \text{ for some } v\in H(v_o)\cap V_G, v' = c_{v}(a_{v \to w}),\\
0 \text{ else}
\end{cases}
\]

It is easy to see that given $w$, there is at most one such $v$ regardless of $G$, 
and exactly those $G$ are deemed responsible which contain the agent choosing at $v$, i.e., for which $v\in V_G$.

Clearly, this account does not take any form of uncertainty into account, and therefore gets neither the `load and shoot' nor the `rock throwing' examples correctly. In those nodes where agent $j$ does not actually have an influence, but they do not know this for sure we argue that they should assume responsibility while the strict causation account assigns zero. This is one of the motivations why we argue that a probabilistic account of causation taking into account uncertainty is more valuable.

\subsection{Variant 1: increasing guaranteed likelihood.}

In this variant we begin by taking uncertainty into account, thus eliminating the very first criticism we had for the benchmark account. This reasoning serves as the very motivation for why we chose to include uncertainty in our model and why we employ a probabilistic account of causation.
The rationale for this variant is to translate the basic idea of the stit approach into a probabilistic context, by sticking to a `certain direct influence' view of responsibility.
That is, backward responsibility can be seen as arising from 
having caused an increase in the guaranteed likelihood of an undesired outcome. This is also in line with the conception of \cite{vallentyne2008bruteluck}.

\paragraph{Guaranteed likelihood, caused increase, backward responsibility.}
We measure the {\em known guaranteed likelihood} of $\eps$ at some node $v\in V$ by the quantity
\begin{align}\label{eq:gamma}
   \gamma(v) &\eqdef  \min_{\sigma\in\Sigma(v)} \min_{\zeta\in Z^\sim(v)} \ell(\eps|v,\sigma,\zeta). 
\end{align}

We measure the {\em caused increase in known guaranteed likelihood} in choosing $a\in A_{v_d}$ at decision node $v\in V_d$ by the difference
\begin{align}\label{eq:Dgamma}
    \Delta\gamma(v,a) &\eqdef \gamma(c_{v}(a)) - \gamma(v).
\end{align}
Note that by definition of $\gamma$, $\Delta\gamma(v,a)\ge 0$ for all $a \in A_{v}$.

To measure $G$'s {\em backward responsibility} regarding $\eps$ in outcome node $w\in V_o$,
in this variant we take their aggregate caused increases over all decisions $a_{v \to w}$ for $v\in V_G$.
\begin{align}
    \R_b^1(w) &\eqdef \sum_{v\in H(w)\cap V_G} \Delta\gamma(v,a_{v \to w}).
\end{align}

\paragraph{Maximum caused increase, forward responsibility.}
Finally, to measure $G$'s {\em forward responsibility} regarding $\eps$ in decision node $v\in V_G$,
we take the maximal possible caused increase,
\begin{align}
    \R_f^1(v) &\eqdef \max_{a\in A_{v}} \Delta\gamma(v,a).
\end{align}

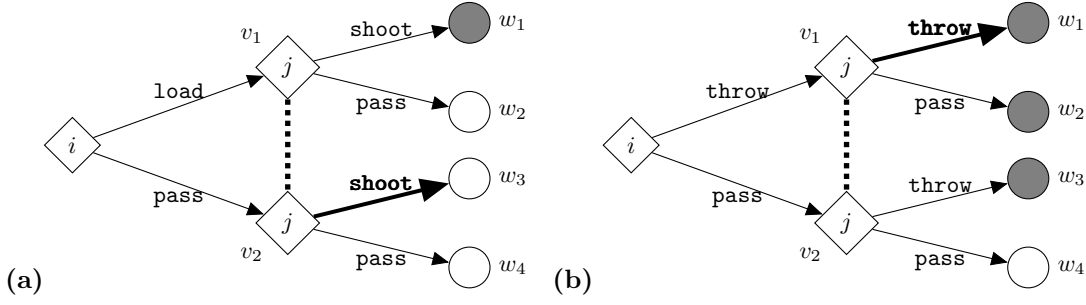
\begin{figure}
\subfigure{
    {\bf (a)}\begin{tikzpicture} [-, scale=.9, every node/.style={scale=.9}, node distance=1cm, decision node/.style={diamond, draw, aspect=1, minimum height= 7mm},  prob node/.style={regular polygon,regular polygon sides=4, draw, minimum height= 1cm},  dummy/.style = {}, good outcome/.style={circle, draw, minimum width=6mm}, bad outcome/.style={circle, draw, fill=gray, minimum width=6mm }, >= triangle 45]
\node[decision node] (1) [] [] {$i$};

\node[dummy] (a) [right of = 1] [] {};
\node[dummy] (a2) [right of = a] [] {};

\node[decision node] (2) [above right =  of a2] [label=above left: $v_1$] {$j$};
\node[decision node] (3) [below right = of a2] [label=below left: $v_2$] {$j$};

\node[dummy] (b) [right of = 2] [] {};
\node[dummy] (b2) [right of = b] [] {};
\node[dummy] (b3) [right of = b2] [] {};

\node[bad outcome] (good2) [above right = .4cm of b2] [label = right: $w_1$]{};
\node[good outcome] (bad2) [below right = .4cm of b2] [label = right: $w_2$] {};

\node[dummy] (c) [right of = 3] [] {};
\node[dummy] (c2) [right of = c] [] {};

\node[good outcome] (bad3) [above right = .4cm of c2] [label = right: $w_3$] {};
\node[good outcome] (good3) [below right = .4cm of c2] [label = right: $w_4$] {};

\path[->] (1) edge node[above] {$\mathtt{load}$}  (2) (1) edge node[below] {$\mathtt{pass}$} (3) (2) edge node[above] {$\mathtt{shoot}$}  (good2) (2) edge node[below] {$\mathtt{pass}$} (bad2) (3) edge node[below] {$\mathtt{pass}$} (good3);
\path[->, line width = 1.5pt](3) edge node[above] {$\pmb{\mathtt{shoot}}$} (bad3);

\draw[line width=2pt, dotted=on] (2) -- (3); 

\end{tikzpicture}}
\subfigure{
    {\bf (b)}
    \begin{tikzpicture} [-, scale=.9, every node/.style={scale=.9}, node distance=1cm, decision node/.style={diamond, draw, aspect=1, minimum height= 7mm},  prob node/.style={regular polygon,regular polygon sides=4, draw, minimum height= 1cm},  dummy/.style = {}, good outcome/.style={circle, draw, minimum width=6mm}, bad outcome/.style={circle, draw, fill=gray, minimum width=6mm }, >= triangle 45]
\node[decision node] (1) [] [] {$i$};

\node[dummy] (a) [right of = 1] [] {};
\node[dummy] (a2) [right of = a] [] {};

\node[decision node] (2) [above right =  of a2] [label=above left: $v_1$] {$j$};
\node[decision node] (3) [below right = of a2] [label=below left: $v_2$] {$j$};

\node[dummy] (b) [right of = 2] [] {};
\node[dummy] (b2) [right of = b] [] {};
\node[dummy] (b3) [right of = b2] [] {};

\node[bad outcome] (good2) [above right = .4cm of b2] [label = right: $w_1$]{};
\node[bad outcome] (bad2) [below right = .4cm of b2] [label = right: $w_2$] {};

\node[dummy] (c) [right of = 3] [] {};
\node[dummy] (c2) [right of = c] [] {};

\node[bad outcome] (bad3) [above right = .4cm of c2] [label = right: $w_3$] {};
\node[good outcome] (good3) [below right = .4cm of c2] [label = right: $w_4$] {};

\path[->] (1) edge node[above] {$\mathtt{throw}$}  (2) (1) edge node[below] {$\mathtt{pass}$} (3) (2) edge node[below] {$\mathtt{pass}$} (bad2)  (3) edge node[below] {$\mathtt{pass}$} (good3) (3) edge node[above] {$\mathtt{throw}$} (bad3);
\path[->, line width = 1.5pt ] (2) edge node[above] {$\pmb{\mathtt{throw}}$}  (good2) ;

\draw[line width=2pt, dotted=on] (2) -- (3); 
\end{tikzpicture}}

\caption{\label{fig:var1and2} Repetition of the paradigmatic example scenarios. Highlighted choices are those for which the influence on responsibility ascription is difficult to capture correctly.
(a) Load and shoot. $\R^1_b(w_3) = 0$, while we argue that it should be non-zero, as agent $j$ must consider the possibility that their action has negative consequences. 
(b) Rock throwing. Similarly, $\R^2_b(w_1) = 0$, while we argue that it should be non-zero.
}
\end{figure}

\paragraph{Application to paradigmatic examples.}
This variant of the responsibility functions fails for the `load and shoot' example. The example is repeated in Fig.~\ref{fig:var1and2}. In node $v_2$, where the gun was not loaded by agent $i$, agent $j$ cannot actually worsen the worst-case likelihood of the undesired outcome $\eps$, as it always stays zero. However, as we argued previously, agent $j$ should nevertheless be assigned responsibility here, as they cannot be sure that their action will not have a negative influence.

This shows the two main drawbacks of this account: it does not sufficiently take information equivalence into account and it is in a sense too ``optimistic'' by allowing agents to assume the best case.
The next variant tries to resolve these issues.

\subsection{Variant 2: increasing worst-case likelihood}

The previous variant, as we discussed, was in a sense `too optimistic' by assigning responsibility only to the degree that the agent did not minimize \emph{guaranteed} likelihood of the undesired outcome. As any ambiguity or knowledge uncertainty with a possible good outcome immediately sets guaranteed likelihood to zero, this can be seen as a focus on the `best-case' view. 
Now we want to take the opposite view by looking at the worst case instead.
That is, in this variant, we assign responsibility to the degree that the agent did not minimize worst-case likelihood of the undesired outcome.
In defining the worst case, however, we assume a group $G$ can plan and commit to optimal future behaviour, 
so that we will look at strategies $\sigma$ rather than actions $a$.

\paragraph{Worst case and minimax likelihoods.}
We define $G$'s {\em worst-case likelihood} of $\eps$ at any node $v\in V$ given some strategy $\sigma\in\Sigma(v)$ by
\begin{align}
    \lambda(v,\sigma) &\eqdef \max_{\zeta\in Z^\sim(v)} \ell(\eps|v,\sigma,\zeta).
\end{align}
$G$'s {\em minimax likelihood} regarding $\eps$ at $v$ is taken to be the smallest achievable worst-case likelihood,
\begin{align}\label{eq:mu}
    \mu(v) &\eqdef \min_{\sigma\in\Sigma(v)} \lambda(v,\sigma)
    = \min_{\sigma\in\Sigma(v)} \max_{\zeta\in Z^\sim(v)} \ell(\eps|v,\sigma,\zeta).
\end{align}

\paragraph{Worst-case increase, backward responsibility.}
We measure $G$'s {\em worst-case increase in minimax likelihood} 
in choosing $a\in A_{v}$ at node $v\in V_G$ by taking the difference in minimax likelihood before and after $a$. Again we assume the worst case, by taking the highest value over all nodes in $v$'s information set.\footnote{Note that while we did already range over all nodes in the information set in the definition of worst-case likelihood, we need to include them again here. We are now comparing the states after an action with the states before, which can evidently differ depending on the decision node in which the action was effected.}
\begin{align}\label{eq:Dmu}
    \Delta\mu(v,a) &\eqdef \max_{v'\sim v}[\mu(c_{v'}(a)) - \mu(v')] \ge 0
\end{align}

Similarly to before, to measure $G$'s {\em backward responsibility} for $\eps$ in node $w$,
we take the aggregate of the worst-case increases in minimax likelihood.
\begin{align}
    \R_b^2(w) &\eqdef \sum_{v\in H(w)\cap V_G} \Delta\mu(v,a_{v\to w})
\end{align}

\paragraph{Maximum caused increase, forward responsibility.}
In analogy to variant 1, to measure $G$'s {\em forward responsibility} regarding $\eps$ in node $v_d\in V_G$,
we take the maximal possible worst-case increase in minimax likelihood, 
\begin{align}
    \R_f^2(v) &\eqdef \max_{a\in A_{v}} \Delta\mu(v,a).
\end{align}

\paragraph{Application to paradigmatic examples, discussion.}
This variant correctly captures the intuition that we should assign responsibility to agent $j$ in the `load and shoot' scenario if they do shoot, even if the gun was not loaded. However, it fails for the `rock throwing' example. The worst-case likelihood of the undesirable outcome when the second agent $j$ has to decide is always one, as there is a scenario in which agent $i$ threw the rock and the undesired outcome will certainly happen. Thus, so is the minimax likelihood and no further increase is possible. 

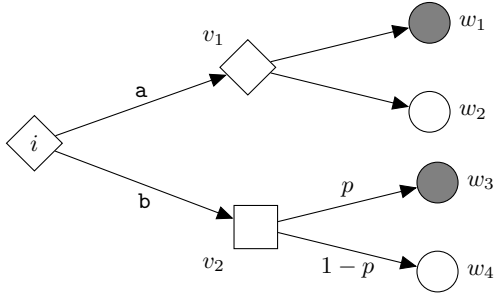
\begin{figure}
\centering
   \begin{tikzpicture} [-, scale=.9, every node/.style={scale=.9}, node distance=1cm, decision node/.style={diamond, draw, aspect=1, minimum height= 7mm},  prob node/.style={regular polygon,regular polygon sides=4, draw, minimum height= .9cm},  dummy/.style = {}, good outcome/.style={circle, draw, minimum width=6mm}, bad outcome/.style={circle, draw, fill=gray, minimum width=6mm }, >= triangle 45]
\node[decision node] (1) [] [] {$i$};

\node[dummy] (a) [right of = 1] [] {};
\node[dummy] (a2) [right of = a] [] {};

\node[decision node] (2) [above right =  of a2] [label=above left: $v_1$] {\ \ \ \ };
\node[prob node] (3) [below right = of a2] [label=below left: $v_2$] {};

\node[dummy] (b) [right of = 2] [] {};
\node[dummy] (b2) [right of = b] [] {};
\node[dummy] (b3) [right of = b2] [] {};

\node[bad outcome] (good2) [above right = .4cm of b2] [label = right: $w_1$]{};
\node[good outcome] (bad2) [below right = .4cm of b2] [label = right: $w_2$] {};

\node[dummy] (c) [right of = 3] [] {};
\node[dummy] (c2) [right of = c] [] {};

\node[bad outcome] (bad3) [above right = .4cm of c2] [label = right: $w_3$] {};
\node[good outcome] (good3) [below right = .4cm of c2] [label = right: $w_4$] {};

\path[->] (1) edge node[above] {$\mathtt{a}$}  (2) (1) edge node[below] {$\mathtt{b}$} (3) (2) edge (bad2) (2) edge (good2) (3) edge node[above] {$p$} (bad3) (3) edge node[below] {$1-p$}   (good3) ;
\end{tikzpicture}
\caption{\label{fig:ambig}
Situation related to ambiguity aversion in which the complementarity of variants 1 and 2 of our responsibility functions can be seen.
The agent must choose between an ambiguous course $\mathtt{a}$ and a risky course $\mathtt{b}$. The ambiguous course leads to a lower responsibility ascription in variant 1 since it does not increase the guaranteed likelihood of a bad outcome. Contrastingly, in variant 2 it is the risky course that leads to a lower responsibility ascription, since this action reduces the minimax likelihood of a bad outcome from 1 to $p$.}
\end{figure}

This variant is in a sense the opposite of variant 1 with respect to its ambiguity attitude. 
To understand their relationship, consider the situation depicted in Fig.~\ref{fig:ambig}.
Variant 1 is ambiguity-affine by favoring ambiguity over risk, while variant 2 is ambiguity-averse, by favoring risk over ambiguity.

In a way, both the previous variant and this one do not sufficiently embrace the complexity arising from uncertainty, but try to reduce the situation to a simple comparison with a best- or worst-case scenario. In the next iteration we aim at improving this.

\subsection{Variant 3: measuring responsibility in terms of influence and risk-taking}

While variants 1 and 2 can be interpreted as measuring the deviation from a single optimal strategy
that minimizes either the guaranteed (best-case) or the worst-case likelihood of the undesired outcome,
our next variant is based on families of scenario-dependent optimal strategies. That is, rather than first reducing to a single scenario and evaluation the choice within this, we evaluate possible strategies in all available scenarios.
In this way, this variant partially manages to avoid being too optimistic or too pessimistic
and thereby captures both the `load and shoot' as well as the `rock throwing' examples correctly. 
The main idea is that backward responsibility arises from taking risks to not avoid an undesirable outcome.

In addition to this new approach to backward responsibility ascription, we also change the way that we treat forward responsibility. In the previous variants, we computed forward responsibility in a certain node as the maximal contribution to backward responsibility that the agent may effect through actions in this node. However, as \citet{vandepoel2011} pointed out, the relation between these two measures of responsibility need not be so straightforward. Therefore, we now employ a separate, independent measure.

\paragraph{Optimum, shortfall, risk, backward responsibility.}
Given a node $v\in V$ and a scenario at $v$, $\zeta\in Z^\sim(v)$, we call
the {\em optimum} $G$ could achieve for avoiding $\eps$ at that node in that scenario the minimum likelihood for $\eps$ over $G$'s strategies at $v$,
\begin{align}
    \omega(v,\zeta) &\eqdef  \min_{\sigma\in\Sigma(v)} \ell(\eps|v,\sigma,\zeta).
\end{align}

So let us measure $G$'s hypothetical {\em shortfall} in avoiding $\eps$ in scenario $\zeta$ 
due to their choice $a\in A_{v}$ at node $v\in V_G$
by the difference in optima
\begin{align}
    \Delta\omega(v,\zeta,a) &\eqdef  
    \omega(c_{{v_\zeta}}(a),\zeta) - \omega(v,\zeta) \ge 0.
\end{align}
Then let the {\em risk taken} by $G$ in choosing $a$ be the maximum shortfall over all scenarios at $v$,
\begin{align}
    \rho(v,a) &\eqdef  \max_{\zeta\in Z^\sim(v)} \Delta\omega(v,\zeta,a).
\end{align}
That is, rather than first reducing the situation to a simple best or worst-case scenario and then comparing strategies within this reduced setting, in this definition we look at strategies for all scenarios.

To measure $G$'s {\em backward responsibility} regarding $\eps$ in node $w\in V_o$,
we again take their aggregate risk taken over all choices they made,
\begin{align}
    \R_b^3(w) &\eqdef  \sum_{v\in H(w)\cap V_G} \rho(v,a_{v\to w}).
\end{align}

\paragraph{Influence, forward responsibility.}
Regarding forward responsibility, we test a different approach than before,
which is simpler but less strongly linked to backward responsibility. We now consider forward responsibility to be strongly related to the concept of influence.
The rationale is that since $G$ does not know which scenario applies, 
they must take into account that their actual influence on the likelihood of $\eps$
might be as large as the maximum of this over all possible scenarios, 
so the larger this value is, the more careful $G$ need to make their choices.

Let us measure $G$'s {\em influence} regarding $\eps$ in scenario $\zeta$ at any node $v\in V$ 
by the range of likelihoods spanned by $G$'s strategies at $v$,
\begin{align}
    \Delta\ell(v,\zeta) &\eqdef \max\{  \ell(\eps|v,\sigma_1,\zeta) - \ell(\eps|v,\sigma_2, \zeta) : \sigma_1, \sigma_2 \in \Sigma(v)\}
\end{align}
To measure $G$'s {\em forward responsibility} regarding $\eps$ at node $v\in V_G$,
we this time simply take their maximum influence over all scenarios at $v$,
\begin{align}
    \R_f^3(v) &\eqdef \max_{\zeta\in Z^\sim(v)} \Delta\ell(v,\zeta).
\end{align}

\paragraph{Application to paradigmatic examples, discussion.}
With this variant of responsibility functions we manage to capture both the `load and shoot' as well as the `rock throwing' examples correctly. 

Let us go through the analysis for the `load and shoot' example in detail. Recall that we are  specifically interested in that path in which agent $i$ did not load the gun and agent $j$ decided to shoot. When agent $j$ has to decide, there are two possible scenarios: the one where agent $i$ did load the gun, let us call this $\zeta_{\mathtt{load}}$, and the one where they did not, $\zeta_{\mathtt{pass}}$. There are also two strategies, shooting $\sigma_{\mathtt{shoot}}$ and not shooting $\sigma_{\mathtt{pass}}$. Then \[\omega(v_2, \zeta_{\mathtt{load}}) = \min_{\sigma\in\Sigma(v_2)} \ell(\eps|v_2,\sigma,\zeta_{\mathtt{load}}) = \ell(\eps|v_2,\sigma_{\mathtt{pass}},\zeta_{\mathtt{load}}) = 0\]
\[\Delta\omega(v_2, \zeta_{\mathtt{load}}, \mathtt{shoot}) = \omega(c_{v_1}(\mathtt{shoot}), \zeta_{\mathtt{load}}) - \omega(v_2, \zeta_{\mathtt{load}}) = \omega(w_1, \zeta_{\mathtt{load}}) - 0 = 1\]
Thus the risk $\rho (v_2, \mathtt{shoot}) = 1$, and the same holds for the backward responsibility \[\R^3_b(w_3, \{j\}) = \sum_{v\in H(w_3)\cap V_j} \rho(v,a_{v \to w_3}) = \rho (v_2, \mathtt{shoot}) = 1 \]
Note that for the computation of the shortfall the scenario $\zeta_{\mathtt{load}}$ is more important than the actual node $v_2$. This is due to the fact that the agent does not know which node they are actually in, so placing emphasis on this node in the reasoning would assume a knowledge the agent does not have.

The analysis for the `rock throwing' example is parallel to this one and left as an easy exercise to the reader.

\begin{figure}
\subfigure{
    {\bf (a)}\begin{tikzpicture} [-, scale=.85, every node/.style={scale=.85}, node distance=1cm, decision node/.style={diamond, draw, aspect=1, minimum height= 7mm},  prob node/.style={regular polygon,regular polygon sides=4, draw, minimum height= .9cm},  dummy/.style = {}, good outcome/.style={circle, draw, minimum width=6mm}, bad outcome/.style={circle, draw, fill=gray, minimum width=6mm }, >= triangle 45]
\node[decision node] (1) [] [] {$\ \ \ $};

\node[dummy] (a) [right of = 1] [] {};
\node[dummy] (a2) [right of = a] [] {};

\node[decision node] (2) [above right =  of a2] [label=above left: $v_1$] {$i$};
\node[decision node] (3) [below right = of a2] [label=below left: $v_2$] {$i$};

\node[dummy] (b) [right of = 2] [] {};
\node[dummy] (b2) [right of = b] [] {};
\node[dummy] (b3) [right of = b2] [] {};

\node[good outcome] (good2) [above right = .4cm of b2] [label = right: $w_1$]{};
\node[bad outcome] (bad2) [below right = .4cm of b2] [label = right: $w_2$] {};

\node[dummy] (c) [right of = 3] [] {};
\node[dummy] (c2) [right of = c] [] {};

\node[bad outcome] (bad3) [above right = .4cm of c2] [label = right: $w_3$] {};
\node[good outcome] (good3) [below right = .4cm of c2] [label = right: $w_4$] {};

\path[->] (1) edge node[above] {$\mathtt{no\ risk} $}  (2) (1) edge node[below] {$\mathtt{cooling}$} (3) ;
\path[->, line width = 1.5pt] (2) edge node[below] {$\pmb{\mathtt{heat\  up\ }} $} (bad2)  (3) edge node[above] {$\pmb{\mathtt{pass} }$} (bad3);
\path[->, line width = 1.7pt, dashed = on] (2) edge node[above] {$\pmb{\mathtt{pass}}$}  (good2) (3) edge node[below] {$\pmb{\mathtt{heat\ up\ }} $} (good3);

\draw[line width=2pt, dotted=on] (2) -- (3); 

\end{tikzpicture}}
\subfigure{
    {\bf (b)}
    \begin{tikzpicture} [-, scale=.85, every node/.style={scale=.85}, node distance=1cm, decision node/.style={diamond, draw, aspect=1, minimum height= 7mm},  prob node/.style={regular polygon,regular polygon sides=4, draw, minimum height= .9cm},  dummy/.style = {}, good outcome/.style={circle, draw, minimum width=6mm}, bad outcome/.style={circle, draw, fill=gray, minimum width=6mm }, >= triangle 45]
\node[decision node] (1) [] [label=above left: $v_1$] {$i$};

\node[dummy] (a) [right of = 1] [] {};

\node[good outcome] (2) [above right =  of a] [label=above right: $w_1$] {};
\node[prob node] (3) [right of = a] [] {};

\node[dummy] (b) [right of = 3] [] {};
\node[dummy] (b2) [right of = b] [] {};

\node[decision node] (4) [above right = .4cm of b2] [label=above left: $v_2$]{$i$};
\node[bad outcome] (bad) [below right = .4cm of b2] [label = right: $w_4$] {};

\node[dummy] (c) [right of = 4] [] {};
\node[dummy] (c2) [right of = c] [] {};

\node[good outcome] (good2) [above right = .4cm of c2] [label = right: $w_2$] {};
\node[bad outcome] (bad2) [below right = .4cm of c2] [label = right: $w_3$] {};

\path[->] (1) edge node[above] {$\mathtt{rescue}$}  (2) (3) edge node[above] {$1-p$} (4) (3) edge node[below] {$p\ \ $} (bad) (4) edge node[below] {$\mathtt{pass}$} (bad2) ;
\path[->, line width = 1.5pt] (1) edge node[below] {$\pmb{\mathtt{pass \ }}$} (3) (4) edge node[above] {$\pmb{\mathtt{rescue}}$} (good2) ;

\end{tikzpicture}}
\caption{\label{fig:candd} Repetition of the paradigmatic example scenarios. Highlighted choices are those for which the influence on responsibility ascription is difficult to capture correctly.
(a) Cooling vs. Heating. With variant 3 of our proposed responsibility functions, agent $j$ will always be assigned responsibility, even if they do select the action that finally leads to the desired outcome (marked by thick dashed arrows). Reversely, with variant 4 agent $j$ will never be assigned responsibility, even if they selec the action that finally leads to the undesired outcome (marked by thick solid arrows).
(b) Hesitation. Using variants 1 and 2, the agents will be assigned partial forward responsibility in node $v_1$ and no backward responsibility in node $w_2$. Variants 3 and 4 capture this correctly by assigning full forward responsibility, and partial backwards responsibility.}
\end{figure}
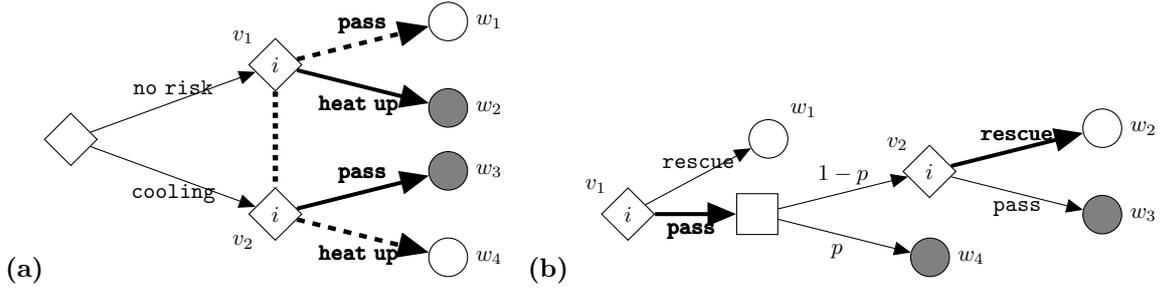

From the paradigmatic examples discussed in section~\ref{sec:paradigmatic}, we have thus far left out the `cooling vs. heating' and `hesitation' examples. These are shown again in Fig.~\ref{fig:candd}. Consider now the `cooling vs. heating' example. The current variant of responsibility functions will assign full backwards- and forwards responsibility in all situations. Backwards responsibility arises from the fact that is impossible for the agents to avoid taking risks. Forward responsibility arises from the realisation that nevertheless, there are scenario-strategy combinations that do lead to the desired outcome. With the next variant we aim at preventing situations like these, where it is impossible for agents to avoid responsibility ascription.

Just like the two variants before, the current variant correctly represents that in the `hesitation' example we do want to assign some backwards responsibility to agent $i$ if they initially hesitated, even if they do in the end rescue the stranger. All three accounts assign responsibility of $p$ in this case. The same holds for outcome node $w_4$ where the agent hesitated and was unlucky.
However, unlike the previous two variants, the current variant also correctly captures that agent $i$ should be assigned full forward responsibility in the initial node $v_1$, as they can ensure any outcome if we take into account the full tree.

\subsection{Variant 4: measuring responsibility in terms of negligence and action-prescribing influence}

In our final variant, we turn the strict assessments of variant 3 in the `cooling vs. heating' coordination example into a more lenient one.
We achieve this by using the actions with the minimum risk as a baseline reference and only `punish' risk-taking exceeding this baseline. The downside of this is that we are consequently faced with responsibility voids. We do not see how any reasonable responsibility ascribing function could be designed that leads neither to voids nor to unavoidable responsibility in coordination examples, so we believe that one has to make a choice between the two.

We also stick with the idea presented in the previous variant of decoupling forwards from backwards responsibility. The issue that we saw previously with the forward responsibility ascription is that agent $i$ was assigned full forward responsibility in the `cooling vs. heating' example, due to the fact that in every scenario their actions can fully influence the outcome. However, this does not give the agent any information on which strategy to select in order to achieve a certain outcome. They cannot {\em willfully} ensure a certain outcome. We aim at reflecting this intuition in our definition.

\paragraph{Risk-minimizing action, negligence, backward responsibility.}
We define the {\em minimal risk} for $G$ in decision node $v\in V_G$ as
\begin{align}\label{eq:alpha}
    \underline\rho(v) &\eqdef \min_{a\in A_{v}} \rho(v,a). 
\end{align}

We now suggest to measure $G$'s degree of {\em negligence} in choosing $a\in A_{v}$ at $v$
by the excess risk with respect to the minimum possible risk,
\begin{align}\label{eq:Drho}
    \Delta\rho(v,a) &\eqdef  \rho(v,a) - \underline\rho(v).
\end{align}

We see that this variant is still sensitive to all scenarios (like variant 3) 
rather than just the best-case or the worst-case. 

Now, to measure $G$'s {\em backward responsibility} regarding $\eps$ in node $w\in V_o$,
we suggest again to take their aggregate negligence over all choices taken,
\begin{align}
    \R_b^4(w) &\eqdef  \sum_{v\in H(w)\cap V_G} \Delta\rho(v,a_{v \to w}).
\end{align}

\paragraph{Consistent influence, forward responsibility.}

Recall from variant 3 that we measured $G$'s {\em influence} regarding $\eps$ in scenario $\zeta$ at any node $v\in V$ by 
\begin{align}
    \Delta\ell(v,\zeta) &\eqdef \max\{  \ell(\eps|v,\sigma_1,\zeta) - \ell(\eps|v,\sigma_2, \zeta) : \sigma_1, \sigma_2 \in \Sigma(v)\},
\end{align}
and took the maximum of this value over all scenarios to arrive at the forward responsibility.

However, this does not tell the agent which action actually leads to the desired or undesired outcomes, as the difference might go in diverging directions for the same strategy pair depending on the scenario.

Therefore, we are interested in whether a certain strategy {\em consistently} leads to high or low likelihoods of $\eps$ starting from a given node $v$, meaning in its peak conclusion that even its minimum is high, or its maximum is low.\footnote{Evidently, intermediate approaches would also be an option. However, as we will see below, for the current application scenario $\R^3$ and $\R^4$ are exactly the same. Therefore, a differentiation between these marginal and other intermediate measures is not currently crucial.} 
To this account, we define
\begin{align}
    \underline{\Delta}\L (v) &\eqdef \max \{\min_{\zeta '\in Z^\sim}\ell(\eps|v,\sigma_1,\zeta') - \max_{\zeta '\in Z^\sim} \ell(\eps|v,\sigma_2, \zeta') : \sigma_1, \sigma_2 \in \Sigma(v)\},
\end{align}
which can be seen to be at most as large as $\R_f^3(v)$ because
\begin{align}
    \underline{\Delta}\L (v) &= \max \{\min_{\zeta '\in Z^\sim}\ell(\eps|v,\sigma_1,\zeta') + \min_{\zeta ''\in Z^\sim} [- \ell(\eps|v,\sigma_2, \zeta'')] : \sigma_1, \sigma_2 \in \Sigma(v)\} \\
     &\le \max \{\min_{\zeta '\in Z^\sim}[\ell(\eps|v,\sigma_1,\zeta') - \ell(\eps|v,\sigma_2, \zeta')] : \sigma_1, \sigma_2 \in \Sigma(v)\} \\
     &\le \max \{\ell(\eps|v,\sigma_1,\zeta') - \ell(\eps|v,\sigma_2, \zeta') : \sigma_1, \sigma_2 \in \Sigma(v), \zeta '\in Z^\sim\} = \R_f^3(v).
\end{align}

In case there is no pair of strategies such that one consistently leads to high likelihoods of $\eps$ while the other consistently leads to low likelihoods of $\eps$, then the value of $\underline{\Delta}\L(v)$ will be negative. Hence we take
\begin{align}
    \R^4_f(v)& \eqdef \max\{0, \underline{\Delta}\L(v) \} \le \R_f^3(v).
\end{align}

\paragraph{Application to paradigmatic examples, discussion.}
As we intended, this variant assigns zero responsibility, both forward- and backward-looking, in the `cooling vs. heating' scenario. It still correctly represents both the `load and shoot' and the `rock throwing' examples as well. Finally, it also captures the intuition that in the `hesitation' example agent $i$ should be assigned full forward responsibility in the initial node $v_1$ and partial backward-looking responsibility in outcome node $w_2$ where they initially hesitated and thus risked the stranger's death.
 
\subsection{Summary of the section}

In this section we presented four different variants of forward- and backward-looking responsibility functions, together with one benchmark variant related to strict causation. We were able to show that initial simple approaches do not correctly capture the previously discussed paradigmatic examples and thus moved on to more refined representations. We saw that in coordination scenarios such as the `cooling vs. heating' example it is inevitable to be faced with either responsibility voids or unavoidable responsibility. We presented functions for each of these options. In the following, we will apply these two variants to the computation of responsibility in our reduced climate change application scenario.

\section{Application}\label{sec:application}

In this section we will apply the well-performing proposed responsibility functions to a stylized version of a real-world scenario and compute quantified responsibility scores.
To this end, we start out by defining the specific example.
This climate decision scenario will not be evaluated `analytically'; rather, we use this as a first example to apply the developed representation and determine responsibility results based on the proposed functions. For this, we will use those responsibility functions that were able to correctly represent the paradigmatic examples, variants 3 and 4.
This example utilizes actual emission values and likelihood ranges resulting from climate modelling. However, it is extremely reduced and simplified and only meant as a proof of concept.

\subsection{Application scenario}

\begin{figure}[t!]
    \centering
   \subfigure{
    \begin{tikzpicture} [-, scale=.85, every node/.style={scale=.85}, node distance=2cm, decision node/.style={diamond, draw, aspect=1, minimum height= 7mm},  prob node/.style={regular polygon,regular polygon sides=4, draw, minimum height= .9cm}, range node/.style={regular polygon, regular polygon sides=3, draw, shape border rotate=90, aspect=1, minimum height= 7mm}, dummy/.style = {}, good outcome/.style={circle, draw, minimum width=6mm}, bad outcome/.style={circle, draw, fill=gray, minimum width=6mm }, >= triangle 45]
\node[range node] (1) [] [] {$\ \ \ $};

\node[dummy] (a2) [right of = 1] [] {$p$ to $q$};
\node[dummy] (a3) [right of = a2] [] {\emph{abbreviates}};

\node[dummy] (b) [below of = 1] [] {};

\end{tikzpicture}}
\subfigure{
   \begin{tikzpicture} [-, scale=.85, every node/.style={scale=.85}, node distance=1cm, decision node/.style={diamond, draw, aspect=1, minimum height= 7mm},  prob node/.style={regular polygon,regular polygon sides=4, draw, minimum height= .9cm},  dummy/.style = {}, good outcome/.style={circle, draw, minimum width=6mm}, bad outcome/.style={circle, draw, fill=gray, minimum width=6mm }, >= triangle 45]
\node[decision node] (1) [] [] {$\ \ \ $};

\node[dummy] (a) [right of = 1] [] {};
\node[dummy] (a2) [right of = a] [] {};

\node[prob node] (2) [above right =  of a2] [label=above left: ] {};
\node[prob node] (3) [below right = of a2] [label=below left: ] {};

\node[dummy] (b) [right of = 2] [] {};
\node[dummy] (b2) [right of = b] [] {};
\node[dummy] (b3) [right of = b2] [] {};

\node[good outcome] (good2) [above right = .4cm of b2] [label = right: ]{};
\node[bad outcome] (bad2) [below right = .4cm of b2] [label = right: ] {};

\node[dummy] (c) [right of = 3] [] {};
\node[dummy] (c2) [right of = c] [] {};

\node[good outcome] (good3) [above right = .4cm of c2] [label = right: ] {};
\node[bad outcome] (bad3) [below right = .4cm of c2] [label = right:] {};

\path[->] (1) edge node[above] {$\mathtt{low} $}  (2) (1) edge node[below] {$\mathtt{high}$} (3) ;
\path[->] (2) edge node[below] {$p $} (bad2)  (2) edge node[above] {$1-p $} (good2);
\path[->] (3) edge node[above] {$1-q$}  (good3) (3) edge node[below] {$q $} (bad3);


\end{tikzpicture}}
    \caption{\label{fig:abbreviation} Abbreviation node used in the depiction of the climate decision scenario. A triangular node with a range of likelihoods noted at is right end is an abbreviation for an ambiguity node, for the range of likelihoods, followed by two probability nodes with the corresponding chances for the undesired outcome.}
\end{figure}
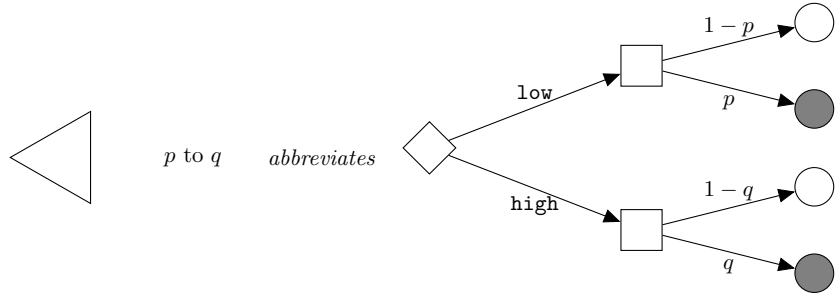

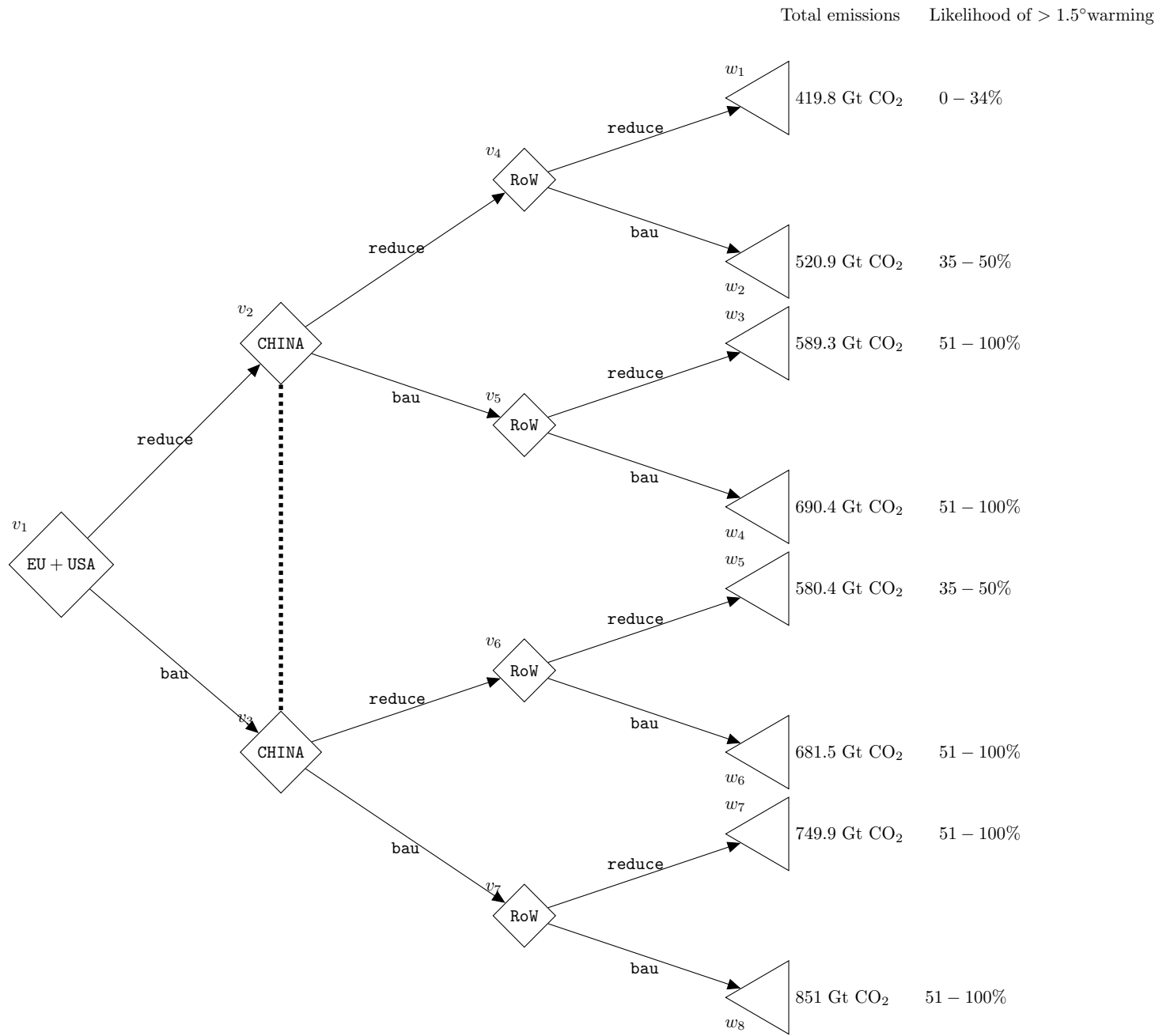
\begin{sidewaysfigure}
\centering
 \begin{tikzpicture} [-, scale=.85, 
 every node/.style={scale=.85}, 
 node distance=1.6cm, 
 decision node/.style={diamond, draw, aspect=1, minimum height= 7mm}, 
 prob node/.style={regular polygon,regular polygon sides=4, draw, minimum height= .9cm}, 
 range node/.style={regular polygon, regular polygon sides=3, draw, shape border rotate=90, aspect=1, minimum height= 7mm}, 
 dummy/.style = {},
 >= triangle 45]


\node[decision node] (v4) [] [label=above left: $v_4$] {$\mathtt{RoW}$};

\node[dummy] (a1) [above of = v4] [] {};

\node[dummy] (a2) [below of = v4] [] {};
\node[dummy] (a3) [below of = a2] [] {};

\node[decision node] (v5) [below of = a3] [label=above left: $v_5$] {$\mathtt{RoW}$};

\node[dummy] (a4) [below of = v5] [] {};
\node[dummy] (a5) [below of = a4] [] {};

\node[decision node] (v6) [below of = a5] [label=above left: $v_6$] {$\mathtt{RoW}$};

\node[dummy] (a6) [below of = v6] [] {};
\node[dummy] (a7) [below of = a6] [] {};

\node[decision node] (v7) [below of = a7] [label=above left: $v_7$] {$\mathtt{RoW}$};

\node[dummy] (a8) [below of = v7] [] {};

\node[dummy] (b1) [right of = a1] [] {};
\node[dummy] (b2) [right of = a2] [] {};
\node[dummy] (b3) [right of = a3] [] {};
\node[dummy] (b4) [right of = a4] [] {};
\node[dummy] (b5) [right of = a5] [] {};
\node[dummy] (b6) [right of = a6] [] {};
\node[dummy] (b7) [right of = a7] [] {};
\node[dummy] (b8) [right of = a8] [] {};

\node[dummy] (bb1) [right of = b1] [] {};
\node[dummy] (bb2) [right of = b2] [] {};
\node[dummy] (bb3) [right of = b3] [] {};
\node[dummy] (bb4) [right of = b4] [] {};
\node[dummy] (bb5) [right of = b5] [] {};
\node[dummy] (bb6) [right of = b6] [] {};
\node[dummy] (bb7) [right of = b7] [] {};
\node[dummy] (bb8) [right of = b8] [] {};

\node[range node] (w1) [right of = bb1] [label=right:$419.8$ Gt CO$_2\qquad 0-34\%$][label=above left: $w_1$] {$\quad$};
\node[range node] (w2) [right of = bb2] [label=right:$520.9$ Gt CO$_2\qquad 35-50\%$] [label=below left: $w_2$] {$\quad$};
\node[range node] (w3) [right of = bb3] [label=right:$589.3$ Gt CO$_2\qquad 51-100\%$] [label=above left: $w_3$] {$\quad$};
\node[range node] (w4) [right of = bb4] [label=right:$690.4$ Gt CO$_2\qquad 51-100\%$] [label=below left: $w_4$] {$\quad$};
\node[range node] (w5) [right of = bb5] [label=right:$580.4$ Gt CO$_2\qquad 35-50\%$] [label=above left: $w_5$] {$\quad$};
\node[range node] (w6) [right of = bb6] [label=right:$681.5$ Gt CO$_2\qquad 51-100\%$] [label=below left: $w_6$] {$\quad$};
\node[range node] (w7) [right of = bb7] [label=right:$749.9$ Gt CO$_2\qquad 51-100\%$] [label=above left: $w_7$] {$\quad$};
\node[range node] (w8) [right of = bb8] [label=right:$851$ Gt CO$_2\qquad 51-100\%$] [label=below left: $w_8$] {$\quad$};

\node[dummy] (dum) [above of = w1] [label = right: Total emissions $\quad$ Likelihood of $> 1.5$\degree warming] {};

\node[dummy] (c3) [left of = a3] [] {};
\node[dummy] (c6) [left of = a6] [] {};

\node[dummy] (cc3) [left of = c3] [] {};
\node[dummy] (cc6) [left of = c6] [] {};

\node[decision node] (v2) [left of = cc3] [label=above left: $v_2$] {$\mathtt{CHINA}$};
\node[decision node] (v3) [left of = cc6] [label=above left: $v_3$] {$\mathtt{CHINA}$};

\node[dummy] (d4) [left of = a4] [] {};
\node[dummy] (dd4) [left of = d4] [] {};
\node[dummy] (ddd4) [left of = dd4] [] {};
\node[dummy] (dddd4) [left of = ddd4] [] {};
\node[dummy] (ddddd4) [left of = dddd4] [] {};

\node[decision node] (v1) [below left of = ddddd4] [label= above left: $v_1$] {$\mathtt{EU+USA}$};


\path[->] (v4) edge node[above] {$\mathtt{reduce} \ \ \ $} (w1) (v4) edge node[below] {$\mathtt{bau}$} (w2); 
\path[->] (v5) edge node[above] {$\mathtt{reduce} \ \ \ $} (w3) (v5) edge node[below] {$\mathtt{bau}$} (w4); 
\path[->] (v6) edge node[above] {$\mathtt{reduce} \ \ \ $} (w5) (v6) edge node[below] {$\mathtt{bau}$} (w6); 
\path[->] (v7) edge node[above] {$\mathtt{reduce} \ \ \ $} (w7) (v7) edge node[below] {$\mathtt{bau}$} (w8); 

\path[->] (v2) edge node[above] {$\mathtt{reduce} \ \ \ $} (v4) (v2) edge node[below] {$\mathtt{bau}$} (v5); 
\path[->] (v3) edge node[above] {$\mathtt{reduce} \ \ \ $} (v6) (v3) edge node[below] {$\mathtt{bau}$} (v7); 

\path[->] (v1) edge node[above] {$\mathtt{reduce} \ \ \ $} (v2) (v1) edge node[below] {$\mathtt{bau}$} (v3); 

\draw[line width=2pt, dotted=on] (v2) -- (v3);

\end{tikzpicture}
\caption{\label{fig:climate}
Reduced version of a decision problem related to the climate crisis. First the EU+USA, agent $i$, get to choose between high or low emissions, then CHINA, agent $j$, gets to make the same choice (adjusted to their emission level), without previously knowing about the other agent's selection. Finally the rest of the world, RoW or agent $k$, makes the same choice but knowing about the previous' agents choices. Consequently, undesired climate tipping occurs within a known range of probabilities.}
\end{sidewaysfigure}

The application scenario within which we analyse the results of our representation is depicted graphically in Fig.~\ref{fig:climate}. We use an abbreviation node for outcomes in a known probability range for a more succinct representation. See Fig.~\ref{fig:abbreviation} for the details of this abbreviation. 
 This example showcases a reduced climate change decision making scenario:
Two agents, $i = \mathtt {EU+ USA} $ and $j = \mathtt{China}$, consecutively get to decide on high or low emissions, depending on their respective emission standards.
Agent $i$ obviously does not know yet what the other will decide. However, also agent $j$ cannot be sure what $i$ will do, as even if countries sign, say, the Paris agreement, it remains unclear whether they will actually follow through with the corresponding actions. Therefore, we assume that the second agent $j$ does not know about agent $i$'s decision.
Afterwards a third agent $k= \mathtt{RoW}$, the rest of the world, faces the same decision, but already knowing what the previous agents selected.
These groups of countries were selected to represent the `early-onset industrialised countries' in contrast to a `growing economy'. China, Europe and the United States combined emitted more than half of country specific CO$_2$ emissions in 2017 \citep{muntean2018}. Thus, the rest of the world was arbitrarily grouped into one for this reduced example. No responsibility ascription will be computed for this third agent. 

The example is set from the beginning of the year 2018 until the end of the year 2040. This allows us to use the statements made in the 2018 IPCC report regarding carbon budgets, that were computed starting from 2018  \citep{ipcc1.5spm}. Setting the end-point at 2040 also allows us to use the given likelihoods of warming, that are more precise in a multi-decadal rather than a longer time scale, while also ensuring that with a `high' emissions strategy the carbon budget would be depleted. Setting the starting time to 2018 evidently sets aside any kind of historical responsibility. For a realistic evaluation of responsibilities in the climate crisis, however, such a historical perspective should not be neglected.

The undesired outcome is a warming of the global mean surface air temperature bigger or equal to 1.5\degree C above pre-industrial levels.
Fig.~\ref{fig:budgetlikely} shows the relations between the remaining carbon emissions starting in 2018 and the likelihood of the undesired outcome as they were assessed in 2017/18.

\begin{table}[]
\centering
\begin{tabular}{r | c | c}
      Carbon emissions after 2017  & Likelihood of warming $<$ 1.5\degree C & Likelihood of warming $\geq$ 1.5\degree C \\
     \hline
    0 -- 420 Gt CO$_2$ & 66\% -- 100\% & 0\% -- 34\% \\
    420 -- 580 Gt CO$_2$ & 50\% -- 65\% & 35\% -- 50\% \\
    $>$ 580 Gt CO$_2$ & 0\% -- 49\% & 51\% -- 100\% \\
\end{tabular}
\caption{\label{fig:budgetlikely}Relations between remaining carbon emissions and likelihoods of warming above 1.5\degree C, evaluated in 2017/2018.}
\end{table}
 
`High' emissions will be conceptualised as continuing to emit CO$_2$ at 2017 levels for the 22 years covered by our example. We call this `business as usual' or $\mathtt{bau}$. These levels are taken from the ``2018 report on fossil CO$_2$ emissions of all world countries'' issued by the publications office of the European Union \citep{muntean2018}. This means for $\mathtt{EU+USA}$ emissions of $\sim9$ GtCO$_2$ per year, for $\mathtt{China}$ $\sim11$ GtCO$_2$ per year and for the rest of the world, $\mathtt{RoW}$, $\sim17$ GtCO$_2$ per year.

`Low' emissions will be conceptualised as a reduction such that the total contribution to the global emissions do not exceed the countries' per capita share of the remaining carbon budget for a 66\% chance of staying within 1.5\degree C warming. Again we make things simple by assuming no population growth. We call this strategy $\mathtt{reduce}$. With the population numbers also given in \citet{muntean2018} this amounts to $46.4$ GtCO$_2$ for $\mathtt{EU+USA}$, $83.5$ GtCO$_2$ for $\mathtt{China}$ and $289.9$ GtCO$_2$ for the rest of the world $\mathtt{RoW}$. This results in the total emissions given in Fig.~\ref{fig:emissions} for each strategy.\footnote{The large discrepancy for the three groups between the difference in amounts given for the business as usual strategies and the reduction strategies shows the unequal distribution in current emissions.}

\begin{table}[]
\centering
\begin{tabular}{c | c | c | c}
     $\qquad$  & $\qquad$  $\mathtt{EU+USA}$ $\qquad$  & $\qquad$ $\mathtt{China}$ $\qquad$ & $\qquad$  $\mathtt{RoW}$ $\qquad$ \\
     \hline
   $\qquad$  $\sigma = \mathtt{reduce}$ $\qquad$  & 46.4 Gt CO$_2$ & 83.5 Gt CO$_2$ & 289.9 Gt CO$_2$ \\
    $\sigma = \mathtt{bau}$ & 207 Gt CO$_2$ & 253 Gt CO$_2$ & 391 Gt CO$_2$  \\
\end{tabular}
\caption{\label{fig:emissions}Emissions per agent for both strategies, over the whole period of time covered by the example.}
\end{table}

\subsection{Evaluation}

Now we compute the forward- and backward-looking responsibilities for $\mathtt{CHINA}$ and $\mathtt{EU+USA}$ in the described example using variants $\R^3$ and $\R^4$ of the proposed responsibility functions, those variants that managed to get the analytically evaluated paradigmatic examples right. It turns out that in this example, variants 3 and 4 are completely unanimous in their ascription.

As we used abbreviation nodes to make the representation of the tree more succinct, we did not name every single outcome node. However, as we consider that actions not under the influence of the agent do not affect responsibility ascription, the backward-looking responsibility will actually be the same for all four outcome nodes contained in one abbreviation node. The name we gave the abbreviation node in Fig. \ref{fig:climate} will be used to talk about those outcome nodes. Thus $w_1$ actually stands for four outcome nodes $\{ w_{1,1}, w_{1,2}, w_{1,3}, w_{1,4}\}$, and a statement such as $\R^4_b(w_1) = x$ is short for $R^4_b(w) = x \ \forall w\in w_1=\{ w_{1,1}, w_{1,2}, w_{1,3}, w_{1,4}\}$.\\
The resulting responsibility scores are summarised in table~\ref{tab:results}. The detailed computation can be found in the Appendix.

It was visible that only in case every agent selects `low' emissions is it possible to keep the likelihood of undesired heating below 34\%, while a combination of China and one other agent selecting `low' emissions with the third agent selecting `high' emissions leads to a likelihood of 35--50\% of undesired warming. All other combinations, that is, all situations in which more agents selected high emissions, lead to a chance of the undesired warming which is greater than 50\%. In particular, as soon as $\mathtt{China}$ selects `high' emissions, the likelihood for undesired warming is above 50\%.

Within this scenario, we computed the backward- and forward-looking responsibilities of $\mathtt{EU+USA}$ as well as $\mathtt{China}$, and were able to show that if they select `low' emissions, neither agent will carry backward-looking responsibility. However, if they select `high' emissions both agents carry backward-looking responsibility $\geq 0.5$, with the score being slightly higher for China, due to its higher impact. Again, recall that we artificially set our starting point to 2018 and disregarded any historical considerations. Forward-looking responsibility in this case coincided completely with maximal backward-looking one.

This shows that, even though China is the only agent who can `see to it that' the highest chance for the undesired outcome obtains by selecting `high' emissions, it is not the only agent who is assigned responsibility. Additionally, despite no one being able to single-handedly `see to it that' the lowest chance for the undesirable outcome obtains, we are still able to assign responsibility. 

As a more interesting comparison, using a NESS account of causation would also result in every agent who selects `high' emissions being assigned backward-looking responsibility, albeit without discrimination regarding the likelihood ranges that their action entails. Consider for example the situation in which $\mathtt{EU+USA}$ and $\mathtt{China}$ select `high' emissions, while the rest of the world selects `low' emissions. In this case, $\mathtt{EU+USA}$ is not a NESS-cause for the likelihood of warming being above 50\%. If they selected `low' emissions the same likelihood range would obtain, while if China selected `low' emissions, a lower likelihood range would obtain. However, going back to a NESS account of causation for undesired warming, this distinction would have to be disregarded. This is because the undesired outcome still lies along the same path of further events, despite being less likely.

\bgroup
\def\arraystretch{2.5}
\begin{table}[]
    \centering
    \begin{tabular}{m{1cm} l | c C{1.5cm} | c C{1.5cm} |}
    \cline{3-6}
    \ & \ & \multicolumn{2}{c|}{ \bfseries{$\mathtt{EU+USA}$} } & \multicolumn{2}{c|}{\bfseries{$\mathtt{CHINA}$}} \\ 
    \hline
    \multicolumn{1}{|l}{\multirow{2}{*}{$\R_b$} }& $\sigma = \mathtt{reduce} $ & \multicolumn{1}{c|}{$v \in \{w_1, \ldots, w_4\}$} & \textbf{0 \%} &\multicolumn{1}{c|}{ $v\in\{w_1, w_2, w_5, w_6\}$} & \textbf{0 \%} \\
    \cline{2-6}
    \multicolumn{1}{|l}{} & $\sigma = \mathtt{bau} $ & \multicolumn{1}{c|}{$v\in \{w_5, \ldots, w_8\}$}  & \textbf{50 \%} & \multicolumn{1}{c|}{ $v\in\{w_3, w_4, w_7, w_8\}$} & \textbf{66 \%}\\
    \hline
    \multicolumn{1}{|l}{\multirow{2}{*}{$\R_f$}} &  & \multicolumn{2}{c|}{\multirow{2}{*}{\bfseries{50 \%}}} & \multicolumn{1}{c|}{$v=v_2$} & \textbf{66 \%} \\
    \cline{5-6}
    \multicolumn{1}{|l}{}&  &  & \ & \multicolumn{1}{c|}{$v=v_3$} & \textbf{66 \%} \\
    \hline
    \end{tabular}
    \caption{Responsibility scores for both agents in the described example when evaluated at node $v$. There is no need to differentiate between $\R^3$ and $\R^4$, as in the current example the two give exactly the same results.}
    \label{tab:results}
\end{table}
\egroup

\section{Conclusion and Outlook}\label{sec:conclusion}


We aimed at providing a representation of degrees of responsibility in complex situations, with specifically the current climate crisis as one application scenario in mind. Complicating features in the relevant decision scenarios are multi-agent interactions, uncertainty, both probabilistic as well as ambiguous, and a temporal progression between decisions.

\paragraph{Framework.} 
We provided a framework to represent all these features of the relevant decision scenarios. The framework we used is based on extensive-form games, as these provide a tool for representing multi-agent interactive decision scenarios over time. We added specific nodes to represent probabilistic and ambiguous uncertainty about the future, and an equivalence relation to represent agents' information uncertainty. A universal `ethical desirability' assessment designates a subset of outcome nodes as undesirable.

\paragraph{Paradigmatic examples.}
Afterwards, we represented a set of example scenarios in this framework. Three of the scenarios represent two-stage interactions of two agents, with the second unaware of the first's decision. In the first example (`load and shoot') the second agent does not know whether each of their actions will lead to a \emph{good} outcome, or if their actions select between {\em good} and {\em bad} outcomes. Due to this uncertainty, we argue that the agent must take into account that their action has an effect and should carry responsibility. The second example (`rock throwing') is parallel, but this time the second agent does not know whether all of their actions lead to a {\em bad} outcome or if they can differentiate between {\em good} and {\em bad} outcomes. Our argument regarding responsibility ascription is the same as before. The third example (`cooling vs. heating') is a coordination game, where the second agent knows for sure that their actions can differentiate between {\em good} and {\em bad} outcomes, however, they do not know which actions leads where. This scenario is hard to analyse, as it is impossible to eschew both responsibility voids and unavoidable responsibility. Finally the last example (`hesitation') is one of a repeated action by the same agent with a probabilistic uncertainty node determining whether the agent `is allowed' a second choice. We argue that repeated actions towards a {\em bad} outcome, or away from a {\em good} one, should lead to higher responsibility scores.

\paragraph{Candidate functions.}
Subsequently, we defined candidate responsibility functions within the previously presented framework. In contrast to existing representations in the area of formal ethics we took a probabilistic approach to causation. 

As a benchmark variant we employed a representation of responsibility related to the most basic view of strict causation. As this variant takes no uncertainty into account, it gets neither of the paradigmatic examples right.
Next we continued with variants using the aforementioned probabilistic account of causation and taking uncertainty into account. In a first variant ($\R_{f/b}^1$) we assumed the best case scenario, and evaluated from there. This was insufficient for the `load and shoot' example, where in the best case scenario the second agent has no influence but in the other scenario they do. 
We managed to correctly represent the `load and shoot' example by taking the opposite approach and assuming the worst case scenario in a second variant ($\R_{f/b}^2$). The prescribed action rationale behind this function can be seen as `avoiding the worst' (in order to avoid carrying responsibility), rather than optimising for the best. However, this approach fails for the `rock throwing' example, with parallel reasoning as before.

Both of the above functions reduce the conceptual complexity of responsibility ascription by relying on comparisons between the action taken at a specific node and an assumed 'baseline case' (specifying both a scenario and the desirable action within this scenario). In order to surpass this limitation, we 
subsequently compare agents' \emph{strategies}, that is, full plans of action for all future decisions to be taken, within \emph{families of scenarios}. 
In the next variant ($\R_{f/b}^3$) we assigned responsibility whenever a hypothetical minimum was not reached. This correctly captured both the `load and shoot' as well as the `rock throwing' examples, and prevented responsibility voids in the coordination game `cooling vs. heating'. However, in this last example it also resulted in $i$ being assigned responsibility no matter what their action was. In the final variant we therefore set a group's best option as the baseline to be compared to ($\R_{f/b}^4$). This still correctly captured the first two examples and avoided $i$ being always to some extent responsible in the coordination example. However, it re-introduced voids that were absent with the preceding function. Both of these functions also capture the last example correctly, and together manage to represent either approach to the coordination example that a modeller might want to take.

\paragraph{Application}
Finally, the last two responsibility variants, those that captured the paradigmatic examples correctly, were applied to a reduced scenario inspired by real-life decision making.
The scenario we examined was one where three agents, EU+USA, China and the rest of the world, successively decided on high or low emissions in the beginning of the year 2018. Outcome nodes were set in the year 2040. High emissions were implemented as continued emissions at 2017 levels for the 22 years covered by the scenario, while low emissions were implemented as a reduction in line with limiting total emissions to the cap given by the 2018 IPCC report for a 66\% chance of staying below 1.5\degree C warming \citep{ipcc1.5spm}. The undesired outcome was a warming of earth's surface temperature higher than 1.5\degree C above pre-industrial levels. The corresponding likelihood ranges were again taken from the 2018 IPCC report.
We computed backward- and forward-looking responsibility scores for EU+USA and China. It turned out that the two variants of the responsibility function that we used gave the same results in this example. We also showed that while the agent who has a higher total influence carries a slightly higher forward- as well as backward-looking responsibility, this does not suffice to relieve the other agent from their responsibility, as is sometimes argued.

Overall, we were able to provide a proof of concept that the suggested responsibility functions are applicable to real-world decision scenarios and provide sufficiently differentiated responsibility scores. Specifically for an application to climate action, likelihoods of outcomes are a significant informant for decision making and this should be reflected in responsibility ascription.


\paragraph{Conclusion.}
In conclusion, we were able to provide a framework for representing decision scenarios which is able to capture all relevant features for responsibility ascription in complex interactive situations such as collective action with respect to climate change. Within this framework we defined a set of possible responsibility functions, giving a quantitative measure based on a probabilistic view of causation. We discussed a set of paradigmatic example scenarios to evaluate the proposed functions and saw that there are situations in which it is impossible to avoid both responsibility voids as well as unavoidable responsibility for the agents. In those cases, we need to select either always assigning responsibility, or to never do. Aside from this selection we showed that two of our proposed responsibility functions, one for each of the mentioned choices, correctly captures all the discussed examples. We applied those functions to a reduced real-world application scenario based on international climate action. We were able to show that our functions provide differentiating responsibility scores in this case, both forward- and backward-looking, and that our approach provides a more fine-grained analysis than an account using, for example, NESS-causation would be able to.

\paragraph{Outlook.}

Two central continuations of the work presented here immediately come to mind.
First of all, we might wish to get a better understanding of which responsibility functions are better suited in which situations. In the current paper we did this by analytically analysing example scenarios and comparing proposed functions by asking whether their evaluation of these scenarios coincides with the previously established one. This can be generalised by abstracting away from concrete application scenarios to universal desirable properties. One example could be an equivalence requirement for decisions taken in the same information set, which would cover both the `load and shoot' as well as the `rock throwing' example. Using these desirable properties in an axiomatic method, ideally the space of possible responsibility functions could be characterised.

Second, clearly the reduced application example presented in section~\ref{sec:application} leads one to consider a large-scale representation of climate relevant decisions of various agents and a subsequent computation of applicable responsibility scores.
In line with this, possibly an evaluation and reduction of the complexity of our proposed functions will become necessary. As our calculations often employ minimal or maximal likelihoods of the undesired outcome the calculation will quickly require many steps if we take into account every action by every agent. However, applying heuristics concerning what the worst and best scenarios will be or means of reducing the underlying tree, should allow for the application of these functions to actual real world decision problems currently at hand.

\paragraph*{Acknowledgements.}
We thank 
Markus Brill,  
Jonas Israel,
Rupert Klein, 
Ulrike Kornek, 
Martin Sch\"onfeld, 
two anonymous referees of an earlier version of the manuscript,
the rest of the ALGO group at Technical University Berlin, 
the copan collaboration at the Potsdam Institute for Climate Impact Research
and the participants of the Formal Ethics workshop in Ghent 2019 for fruitful discussions and 
comments.

This work was funded by the Deutsche Forschungsgemeinschaft (DFG, German Research Foundation) under Germany's Excellence Strategy – The Berlin Mathematics Research Center MATH+ (EXC-2046/1, project ID: 390685689).

\paragraph{Declarations of interest.} None.

\bibliographystyle{abbrvnat}
\bibliography{main}

\section*{Appendix}

\subsection*{Computations of the Responsibility Scores}

Now we will give the details of the computations of the responsibility assignments, both forward- and backward-looking for both agents considered in the example.
Recall the notice on abbreviations of outcome nodes.
As we used abbreviation nodes to make the representation of the tree more succinct, we did not name every single outcome node. However, as we consider that actions not under the influence of the agent do not affect responsibility ascription, the backwards-looking responsibility will actually be the same for all four outcome nodes contained in one abbreviation node. The name we gave the abbreviation node will be used to talk about those outcome nodes. Thus $w_1$ actually stands for four outcome nodes $\{ w_{1,1}, w_{1,2}, w_{1,3}, w_{1,4}\}$, and a statement such as $\R^4_b(w_1) = x$ is short for $R^4_b(w) = x \ \forall w\in w_1=\{ w_{1,1}, w_{1,2}, w_{1,3}, w_{1,4}\}$. 
Additionally we will use the notations $\sigma_{red}$ or $\sigma_{bau}$ for those strategies where the agent selects reduced or business as usual emissions, respectively. Similarly, $\zeta_{red, bau, low}$ stands for that scenario where the other agents select reduced emissions, business as usual emissions and where the ambiguity node is followed by the lower probability branch, in that order.\footnote{Note that, depending on the agent under consideration $\zeta{red, bau, low}$ could mean that $\mathtt{EU+USA}$ selects $\mathtt{reduce}$d emissions or that $\mathtt{CHINA}$ selects $\mathtt{reduce}$d emissions, then $\mathtt{RoW}$ selects $\mathtt{bau}$ emissions and the ambiguity node is followed by the lower probability branch.} 

Before we start, recall the definitions of the responsibility functions. Let a tree $\T$, event $\eps$ and group $G$ be given.

\paragraph{Variant 3, backward responsibility.}
Given a node $v\in V$ and a scenario at $v$, $\zeta\in Z^\sim(v)$, we call
the {\em optimum} $G$ could achieve for avoiding $\eps$ at that node in that scenario the minimum likelihood for $\eps$ over $G$'s strategies at $v$,
\begin{align*}
    \omega(v,\zeta) &\eqdef  \min_{\sigma\in\Sigma(v)} \ell(\eps|v,\sigma,\zeta).
\end{align*}

So we measure $G$'s hypothetical {\em shortfall} in avoiding $\eps$ in scenario $\zeta$ 
due to their choice $a\in A_{v_d}$ at node $v_d\in V_G$
by the difference in optima
\begin{align*}
    \Delta\omega(v_d,\zeta,a) &\eqdef  
    \omega(c_{{v_\zeta}}(a),\zeta) - \omega(v_d,\zeta) \ge 0.
\end{align*}
Then we defined the {\em risk taken} by $G$ in choosing $a$ to be the maximum shortfall over all scenarios at $v_d$,
\begin{align*}
    \rho(v_d,a) &\eqdef  \max_{\zeta\in Z^\sim(v_d)} \Delta\omega(v_d,\zeta,a).
\end{align*}

To measure $G$'s {\em backward responsibility} regarding $\eps$ in node $v_o\in V_o$,
we took their aggregate risk taken over all choices they made,
\begin{align*}
    \R_b^3(v_o) &\eqdef  \sum_{v_d\in H(v_o)\cap V_G} \rho(v_d,a_{v_d\to v_o}).
\end{align*}

\paragraph{Variant 3, forward responsibility.}

We measure $G$'s {\em influence} regarding $\eps$ in scenario $\zeta$ at any node $v\in V$ 
by the range of likelihoods spanned by $G$'s strategies at $v$,
\begin{align*}
    \Delta\ell(v,\zeta) &\eqdef \max\{  \ell(\eps|v,\sigma_1,\zeta) - \ell(\eps|v,\sigma_2, \zeta) : \sigma_1, \sigma_2 \in \Sigma(v)\}
\end{align*}
To measure $G$'s {\em forward responsibility} regarding $\eps$ at node $v_d\in V_G$,
we then took their maximum influence over all scenarios at $v_d$,
\begin{align*}
    \R_f^3(v_d) &\eqdef \max_{\zeta\in Z^\sim(v_d)} \Delta\ell(v_d,\zeta).
\end{align*}

\paragraph{Variant 4, backward responsibility.}
We define the {\em minimal risk} and set of {\em risk-minimizing actions} of $G$ in decision node $v_d\in V_G$ as
\begin{align*}\label{eq:alpha}
    \underline\rho(v_d) &\eqdef \min_{a\in A_{v_d}} \rho(v_d,a). 
\end{align*}

We now measure $G$'s degree of {\em negligence} in choosing $a\in A_{v_d}$ at $v_d$
by the excess risk with respect to the minimum possible risk,
\begin{align*}
    \Delta\rho(v_d,a) &\eqdef  \rho(v_d,a) - \underline\rho(v_d).
\end{align*}

Now, to measure $G$'s {\em backward responsibility} regarding $\eps$ in node $v_o\in V_o$,
we took their aggregate negligence over all choices taken,
\begin{align}
    \R_b^4(v_o) &\eqdef  \sum_{v_d\in H(v)\cap V_G} \Delta\rho(v_d,a_{v_d \to v_o}).
\end{align}

\paragraph{Variant 4, forward responsibility.}

Here, we were interested in whether a certain strategy {\em consistently} leads to high or low likelihoods of $\eps$ starting from a given node $v$, meaning even it's minimum is high, or it's maximum is low. 

\begin{align*}
    \underline{\Delta}\L (v) &\eqdef \max \{\min_{\zeta '\in Z^\sim}\ell(\eps|v,\sigma_1,\zeta') - \max_{\zeta '\in Z^\sim} \ell(\eps|v,\sigma_2, \zeta') : \sigma_1, \sigma_2 \in \Sigma(v)\}
\end{align*}

To avoid a negative value we take
\begin{align}
    \R^4_f(v)& \eqdef \max\{0, \underline{\Delta}\L(v) \} \le \R_f^3(v).
\end{align}

\paragraph{Backward-looking responsibility, EU+USA.}
Now we begin with computing the actual scores in the described example.\\
Let $G = \{ \mathtt{EU+USA} \}$, $\sigma^* = \mathtt{reduce}$, $v=v_1$.
Then $\min\limits_{\sigma \in \Sigma(v)} \ell(\eps| v, \sigma, \zeta) = \ell(\eps|v, \sigma_{red}, \zeta) \quad \forall \zeta \in Z$, thus

\[ \omega (v, \zeta) = \min_{\sigma \in \Sigma(v)} \ell(\eps|v, \sigma, \zeta) = \begin{cases}
0\%, \ \ \qquad \qquad \qquad \zeta_{red, red, low}\\
34\%,\qquad \qquad \qquad \zeta_{red, red, high}\\
35\%, \quad \text{ for }\zeta = \quad \zeta_{red, bau, low}\\
50\%, \qquad \qquad \qquad \zeta_{red, bau, high} \\
51\%, \qquad \qquad \qquad \zeta_{bau, red, low} \zeta_{bau, bau, low}\\
100\% \qquad \qquad \qquad \zeta_{bau, red, high} \zeta_{bau, bau, high}
\end{cases} \] 

And equivalently \[ \omega (c_{v_{\zeta}}(\mathtt{reduce}), \zeta) = \min_{\sigma\in \Sigma(v_2)} \ell(\eps| v_2, \sigma, \zeta) = \begin{cases}
0\%, \ \ \qquad \qquad \qquad \zeta_{red, red, low}\\
34\%,\qquad \qquad \qquad \zeta_{red, red, high}\\
35\%, \quad \text{ for }\zeta = \quad \zeta_{red, bau, low}\\
50\%, \qquad \qquad \qquad \zeta_{red, bau, high} \\
51\%, \qquad \qquad \qquad \zeta_{bau, red, low} \zeta_{bau, bau, low}\\
100\% \qquad \qquad \qquad \zeta_{bau, red, high} \zeta_{bau, bau, high}
\end{cases} \]

\begin{align*}
    \text{Thus } \ \omega(c_{v_{\zeta}}(\mathtt{reduce}), \zeta) &= \omega (v, \zeta) \qquad \forall \zeta \in Z \\
    \Rightarrow \rho(\mathtt{EU+USA}, v, \mathtt{reduce}) &= 0 \\
    \Rightarrow \underline \rho(\mathtt{EU+USA}, v_1) &= 0 \\
    \Rightarrow \R^3_b(w) &= R^4_b(w) \qquad \forall w \in V_o\\
    \text{and } \R^3_b(\mathtt{EU+USA}, w) &= \R^4_b(\mathtt{EU+USA}, w) = 0 \qquad \text{ for } w\in \{w_1, \ldots, w_4\}
\end{align*}

Now let $\sigma^* = \mathtt{bau}$. $\omega(v, \zeta)$ remains unchanged. However,

\[ \omega (c_{v_{\zeta}}(\mathtt{bau}), \zeta) = \min_{\sigma\in \Sigma(v_3)} \ell(\eps| v_3, \sigma, \zeta) = \begin{cases}
0\%, \ \ \qquad \qquad \qquad -\\
34\%,\qquad \qquad \qquad -\\
35\%, \quad \text{ for }\zeta = \quad \zeta_{red, red, low}\\
50\%, \qquad \qquad \qquad \zeta_{red, red, high} \\
51\%, \qquad \qquad \qquad \zeta_{red, bau, low}, \zeta_{bau, red, low} \zeta_{bau, bau, low}\\
100\% \qquad \qquad \qquad \zeta_{red, bau, high}, \zeta_{bau, red, high} \zeta_{bau, bau, high}
\end{cases} \] 

Thus

\[ \Delta \omega (v_1, \zeta, \mathtt{bau}) = \omega(v_3, \zeta) - \omega(v_1, \zeta) = \begin{cases}

0\%, \ \ \qquad \qquad \quad \zeta_{bau, red, low}, \zeta_{bau, red, high}, \zeta_{bau, bau, low}, \zeta_{bau, bau, high}\\
16\% ,\quad \text{ for }\zeta = \quad \zeta_{red, red, high}, \zeta_{red, bau, low} \\
35\%, \qquad \qquad \qquad \zeta_{red, red, low} \\
50\%, \qquad \qquad \qquad \zeta_{red, bau, high}
\end{cases}  \] 

\begin{align*}
\text{Hence } \  \rho (v_1, \mathtt{bau}) &= \max\limits_{\zeta \in Z} \Delta \omega (v_1, \zeta, \mathtt{bau})= 50\% \\
\text{and } \R^3_b(\mathtt{EU+USA}, w) &= \R^4_b(\mathtt{EU+USA}, w) = 0.5 \quad \text{ for } w\in\{w_5, \ldots, w_8\}. 
\end{align*}

\paragraph{Backward-looking responsibility, China.}
The computation for $\mathtt{CHINA}$ is equivalent to that of $\mathtt{EU+USA}$ with the following results
\begin{align*}
    \R^3_b(\mathtt{CHINA}, w) &= \R^4_b(\mathtt{CHINA}, w) = 0 \quad \text{ for } w\in\{w_1, w_2, w_5, w_6\} \\
    \text{and } \R^3_b(\mathtt{CHINA}, w) &= \R^4_b(\mathtt{CHINA}, w) = 0.66 \quad \text{ for } w\in\{w_3, w_4, w_7, w_8\}.
\end{align*}

\paragraph{Forward-looking responsibility, EU+USA.}
Let $G = \mathtt{EU+USA}$ and $v= v_1$. We start out by computing $\R^3_f(v)$.
\[\Delta \ell (v, \zeta) = \max \{ \ell (\eps|v, \sigma_1, \zeta) - \ell(\eps|v, \sigma_2, \zeta) | \sigma_1, \sigma_2 \in \Sigma(v) \} = \]
\[ = \begin{cases}
    0\%, \ \qquad \qquad \qquad \zeta_{bau, red, low}, \zeta_{bau, red, low}, \zeta_{bau, bau, low}, \zeta_{bau, bau, high}\\
    16\%, \qquad \qquad \qquad \zeta_{bau, red, high}, \zeta_{red, bau low} \\
    35\%, \quad \text{ for } \zeta = \quad \zeta_{red, red, low}\\
    50\%, \qquad \qquad \qquad \zeta_{red, bau, high}\\
    66\%, \qquad \qquad \qquad -
\end{cases}\] 

\begin{align*}
\text{Hence } \quad \R^3_f(v) = \max_{\zeta \in Z(v)} \Delta \ell(v_d, \zeta) = 0.5.
\end{align*}
So this value is actually equivalent to maximum backward-looking responsibility in this specific case. Now let us continue to computing $\R^4_f(v)$.
\[ \ell(\eps \mid v, \sigma_{red}, \zeta) - \ell(\eps| v_1, \sigma_{bau}, \zeta ) = \begin{cases}
    0\%, \ \qquad \qquad \qquad \zeta_{bau, bau, low}, \zeta_{bau, bau, high}, \zeta_{bau, red, low}, \zeta_{bau, red, high}\\
    -16\%, \qquad \qquad \qquad \zeta_{red, red, high} , \zeta_{red, bau low} \\
    -35\%, \quad \text{ for } \zeta = \quad \zeta_{red, red, low} \\
    -50\%, \qquad \qquad \qquad \zeta_{red, bau, high} \\
    -66\%, \qquad \qquad \qquad -
\end{cases}\] 
\begin{align*}
\text{And } & \ell(\eps \mid v, \sigma_{bau}, \zeta) - \ell(\eps| v_1, \sigma_{red}, \zeta ) = - [\ell(\eps \mid v, \sigma_{red}, \zeta) - \ell(\eps| v_1, \sigma_{bau}, \zeta ) ].\\
\text{Thus }\ & \min_{\zeta '\in Z^\sim}\ell(\eps|v,\sigma_{red},\zeta') - \max_{\zeta '\in Z^\sim} \ell(\eps|v,\sigma_{bau}, \zeta') = -0.5,\\
\text{while }\ & \min_{\zeta '\in Z^\sim}\ell(\eps|v,\sigma_{bau},\zeta') - \max_{\zeta '\in Z^\sim} \ell(\eps|v,\sigma_{red}, \zeta') = 0.5.\\
\text{Hence }\ & \max\{  \min_{\zeta '\in Z^\sim}\ell(\eps|v,\sigma_1,\zeta') - \max_{\zeta '\in Z^\sim} \ell(\eps|v,\sigma_2, \zeta') | \sigma_1, \sigma_2 \in \Sigma (v)\} = 0.5 = \R ^4_f(\mathtt{EU+USA},v1) 
\end{align*}

\paragraph{Forward-looking responsibility, China.}
Again, the computation for agent $\mathtt{CHINA}$ runs equivalent to that of agent $\mathtt{EU+USA} $, with the following result
\begin{align*}
    \R^3_f(\mathtt{CHINA}, v_2) &= \R^3_f(\mathtt{CHINA}, v_3) = \R^4_f(\mathtt{CHINA}, v_2) = \R^4_f(\mathtt{CHINA}, v_3) = 0.66.
\end{align*}

\end{document}